\documentclass[useAMS,usenatbib,usefloat]{mn2e}
\usepackage{amsmath}
\usepackage{amssymb}
\usepackage{rotating}
\usepackage{tabularx}
\usepackage{ gensymb }

\usepackage[font=footnotesize,labelfont=bf]{caption}

\newcommand{\ugr}{(u-g, g-r)}
\newcommand{\ltf}{\log(\rm T_{\rm eff})}

\begin{document}

\title[\bf New OB stars in the Carina Arm]{\bf New OB star candidates in the Carina Arm around Westerlund 2 from VPHAS+}
\author[M. Mohr-Smith et al]
{
{\parbox{\textwidth}{M. Mohr-Smith,$^{1}$\thanks{E-mail: m.smith10@herts.ac.uk}
J. E. Drew,$^{1}$
G. Barentsen,$^{1}$
N. J. Wright,$^{1}$
R. Napiwotzki,$^{1}$
R. L. M. Corradi,$^{2}$
J. Eisl\"offel,$^{3}$
P. Groot,$^{3}$
V. Kalari,$^{4}$
Q. A. Parker,$^{5, 6, 7}$
R. Raddi,$^{8}$
S. E. Sale,$^{9}$
Y. C. Unruh,$^{10}$
J. S. Vink,$^{4}$ and
R. Wesson $^{11}$\\}
}\\
$^{1}$Centre for Astrophysics Research, Science and Technology Research Institute, University of Hertfordshire, Hatfield AL10 9AB, UK \\
$^{2}$Instituto de Astrofisica de Canarias, E-38200 La Laguna, Tenerife, Spain \\
$^{3}$Afdeling Sterrenkunde, Radboud Universiteit Nijmegen, Faculteit NWI, Postbus 9010, NL-6500 GL Nijmegen, the Netherlands \\
$^{4}$ Armagh Observatory, College Hill, Armagh, BT61 9DG, UK \\
$^{5}$ Department of Physics \& Astronomy, Macquarie University, Sydney, NSW 2109 Australia \\
$^{6}$ Research Centre for Astronomy, Astrophysics and Astrophotonics, Macquarie University, Sydney, NSW 2109 Australia \\
$^{7}$ Australian Astronomical Observatory, PO Box 296, Epping, NSW 1710, Australia \\
$^{8}$ Department of Physics, University of Warwick, Gibbet Hill Road, Coventry CV4 7AL, UK \\
$^{9}$ Rudolf Peierls Centre for Theoretical Physics, Keble Road, Oxford OX1 3NP, UK \\
$^{10}$ Department of Physics, Blackett Laboratory, Imperial College London, Prince Consort Road, London, SW7 2AZ, UK \\
$^{11}$ European Southern Observatory, Alonso de Cordova 3107, Casilla 19001, Santiago, Chile
}

\maketitle

\begin{abstract}

O and early B stars are at the apex of galactic ecology, but in the Milky Way, only a minority of them may yet have been identified. We present the results of a pilot study to select and parametrise OB star candidates in the Southern Galactic plane, down to a limiting magnitude of $g=20$. A 2 square-degree field capturing the Carina Arm around the young massive star cluster, Westerlund 2, is examined.  The confirmed OB stars in this cluster are used to validate our identification method, based on selection from the $(u-g, g-r)$ diagram for the region. Our Markov Chain Monte Carlo fitting method combines VPHAS+ $u, g, r, i$ with published $J, H, K$ photometry in order to derive posterior probability distributions of the stellar parameters $\ltf$ and distance modulus, together with the reddening parameters $A_0$ and $R_V$. The stellar parameters are sufficient to confirm OB status while the reddening parameters are
determined to a precision of $\sigma(A_0)\sim0.09$ and $\sigma(R_V)\sim0.08$.
There are 489 objects that fit well as new OB candidates, earlier than $\sim$B2. This total includes 74 probable massive O stars, 5 likely blue supergiants and 32 reddened subdwarfs. This increases the number of previously known and candidate OB stars in the region by nearly a factor of 10.
Most of the new objects are likely to be at distances between 3 and 6 kpc. We have confirmed the results of previous studies that, at these longer distances, these sight lines require non-standard reddening laws with $3.5<R_V<4$.
\end{abstract}


\section{Introduction}

Stars of spectral type O and early B, more massive than $\sim 8
M_{\odot}$, are massive enough to form collapsing
cores at the end of their nuclear-burning lifetimes \citep[see e.g.][]{Langer2012, Smarttetal2009}. It is widely recognised that
these stars - henceforward OB stars - are an important source of
kinetic energy, driving turbulence and mixing of the interstellar
medium, powered by a range of phenomena (stellar winds, wind-blown bubbles, expanding HII regions and supernova explosions). They are the main source of
ultra-violet radiation in galaxies and, being short-lived
($\lesssim40$ Myr), they are excellent tracers of recent star
formation.

In the Galaxy, clusters containing OB stars and OB associations
have played an important role in tracing spiral arm structure
\citep[e.g.][]{Russeil2003,vallee2008}. The typical scale height
estimated for OB stars, forming in the Galactic disk, is a few 10s of pc
\citep[e.g.][]{Reed2000,garmanyetal1982}, in keeping with estimates of
the scale height for giant molecular clouds, their birth sites
\citep[e.g.][]{StarkandLee2005}. OB stars are usually regarded as forming in clustered environments \citep{ZinneckerandYorke2007} and are less common in the field. However, examples of isolated field O stars are known and the question has arisen as to whether these high-mass stars have formed in situ, perhaps as the result of stochastic sampling of the initial mass function (IMF) as outlined by \citet{ParkerGoodwin2007}, or have been ejected from clusters as runaways \citep[see e.g.][]{Zwartetal2010,Bestenlehner2011}. In the Milky Way $\sim96$\% of known O-type stars have been identified as members of young open clusters, OB associations or as otherwise
kinematically linked to clustered environments \citep{deWitetal2005}.
This leaves up to $\sim4$\% of Galactic O-type stars possibly forming in
isolation. Deep comprehensive searches for OB stars away from clusters have not been undertaken hitherto.

As luminous objects detected to great distances across the
Galactic disk and through substantial obscuration, OB stars have long been recognised as a
highly-suitable means for characterising the spatial variation of
interstellar extinction, in terms of both dust column and extinction law
\citep[e.g.][]{cardellietal1989, fitzpatrickandmassa2007}. This is aided by their relatively simple optical near-infrared (OnIR) spectral energy
distributions (SEDs). It follows from this that the more densely we can map the positions and
extinctions towards these luminous probes, the more high-quality empirical
constraints we can set on the 3-D distribution of dust and dust properties across the Galactic Plane.

Both of the above areas of enquiry will be well served by a deeper, more
comprehensive mapping of the OB stars
in the Milky Way.  Past cataloguing efforts have been limited
to brighter, nearer objects
\citep[e.g][]{garmanyetal1982,Reed2003,Maiz-Apellanizetal2004}. Indeed
the most comprehensive collection so far, `The Catalog of Galactic OB Stars'
\citep{Reed2003} contains $\sim16000$ known or suspected OB stars taken from
across the literature: around 95\% of the entries are brighter than $13^{\rm th}$
magnitude in the visual bands.  Now is the right time to push the
magnitude limit much fainter, to $\sim20^{\rm th}$ magnitude, given the
likely delivery of astrometry to this depth by the Gaia mission from $\sim2017$
onwards \citep[both
  parallaxes {\em and} proper motions, for details on
expected performance see][]{deBruijne2012}.  Efficient, purely
photometric selection of OB stars in the field as well as in clusters
continues to be best undertaken at blue optical wavelengths, where
colour selection via the $Q$ method
\citep[initiated by][]{JohnsonandMorgan1953} is
proven to separate O and early B stars from later type stars.

The practical motivation of this paper is to establish a method of photometric
selection and analysis that can form the basis
for a new homogeneous census of Galactic OB stars as faint as $g \simeq 20$.
Based on a restrained extrapolation of the first results presented
here, we can surmise that a new census
will more than double the numbers known. A suitable source for the
new census will be the VST Photometric H$\alpha$
Survey of the Southern Galactic Plane and Bulge
\citep[VPHAS+][]{drewetal2014}.  VPHAS+ is a deep, uniform,
photometric survey of the entire southern Galactic Plane and Bulge in
broad-band $u$, $g$, $r$, $i$ and narrow-band H$\alpha$ filters on
ESO's VLT Survey Telescope (VST).  The survey footprint includes the
entire southern Galactic Plane within the Galactic latitude range of $|b| <
5^{\circ}$.  The VST's OmegaCam imager provides a full square
degree field of view with very good spatial resolution ($0.2"$ pixels
sample a median seeing of 0.8 -- 1.0 arcsec in the $u/g/r$ bands).

\begin{figure*}
\centering
\includegraphics[width=\textwidth]{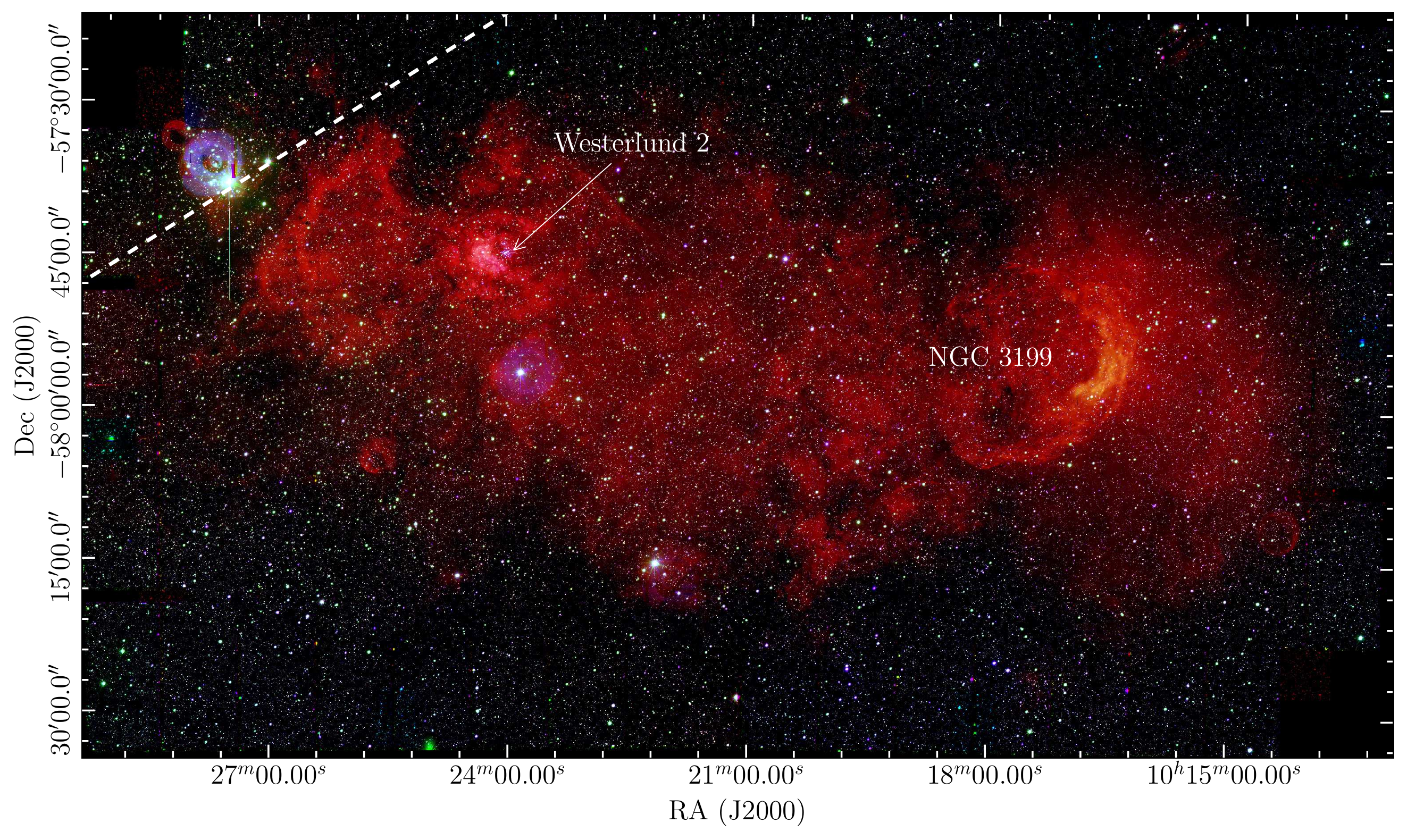}
\caption{RGB image of the $\sim$2 square degree region (H$\alpha$, g, i). This region falls within the star forming complex G283 identified by \protect\cite{rahmanandmurray2010} -- an elliptical region slightly larger than the sky area shown. Westerlund 2 is embedded in the HII region RCW49, while the diffuse nebulae NGC3199 is located to the right (West) as marked. The dashed line traces the Galactic equator.}
\label{fig:colourimage}
\end{figure*}

Here, we present a first study that uses broadband VPHAS+ data to select and parametrize OB stars in a $\sim$2 square-degree area, roughly centred on
$\ell = 284^{\circ}$, $b = -0.7^{\circ}$, in the part of the Plane
containing the young massive cluster, Westerlund 2 (Wd 2), the
larger associated HII region RCW 49, and the diffuse nebula
NGC3199 (see Figure \ref{fig:colourimage}). Previous optical and near-infrared studies on the stellar content of Westerlund 2 have focused on the immediate environment of the cluster itself - a patch of sky ~4 arcmin across - \citep{Moffatetal1991, Ascensoetal2007,  Alvarezetal2013}, while the x-ray study by \cite{Tsujimotoetal2007} focused on an area $\sim17$ arcmin across. Most recently, \cite{huretal2015} have revisited optical photometry of this cluster over a 17.9' x 9.3' footprint.

By tracing $8\mu$m warm-dust emission
\cite{rahmanandmurray2010} have identified this same region as part of
a large star-forming complex (G283).
On the sky, Wd 2 falls close to the Carina Arm tangent direction
\citep[e.g.][]{Russeil2003}: the CO data presented by \citet{Dame2007}
show persuasively that Wd 2 and its environs fall just inside the sky position of the tangent point, but further away. This cluster is estimated to be 1 - 3Myr old \citep{Alvarezetal2013,Ascensoetal2007}. It
contains a large number of spectroscopically-confirmed OB stars,
albeit behind a dust column giving rise to over 6 magnitudes of visual
extinction
\citep{Moffatetal1991,Rauwetal2007,Carraroetal2012,Alvarezetal2013}.
Estimates of the distance to Wd 2 in the literature have varied
enormously, ranging from 2.8~kpc \citep{Ascensoetal2007} up to
$\sim$8~kpc \citep[e.g.][]{Rauwetal2011}.  However, it is not our aim
to enter into this debate.  More important is the likelihood that much
of the scientific gain from VPHAS+ discoveries of OB stars will be in the
domain of visual extinctions of up to 8--10 magnitudes, and distance scales
of 2--10 kpc (according to Galactic longitude).  In this regard, the
field around Wd 2 is highly typical of the task ahead.

A recent study on Wd 2 by \citet{Alvarezetal2013} uses data from
the Hubble Space Telescope (HST) that offers much better spatial
resolution than is achievable from the ground.  This is the only dataset that offers better angular resolution than the new VPHAS+ data analysed
here.  These authors' values of $R_V$ and $A_V$ were derived by
fitting, to 32 individual OB stars in or near Wd 2, reddened model
optical/near-infrared SEDs appropriate for the selected stars'
spectroscopically-confirmed spectral types. The best fits were
computed by seeking the global chi-squared minimum among all plausible
values of $R_V$ and $A_V$ -- resulting in a mean outcome of
$R_V=3.77\pm0.09$ and $A_V=6.51\pm0.38$\,mag combining results from different reddening law prescriptions.  We use a comparison of
our OnIR SED fit results for this same set of OB stars to bench-mark
our method.

This paper is organised as follows.  In Section 2 more details on the
data used for this study are given.  Section 3 is a presentation of
our method, beginning with the updated version of the $Q$ method of
OB star selection that we use, and ending with a description of the Markov Chain Monte Carlo (MCMC) sampling of the posterior distributions of the OnIR SED model fit
parameters. The stage is then set to compare our results for Wd 2
stars with those of \citet{Alvarezetal2013}, in Section 4.  The results of the fits to the final list of 527 new OB candidates drawn
from across the full 2 square degrees are presented in Section 5.
This is followed by a discussion of the results in Section 6, in which we consider the extinction trends revealed in this region, and draw attention to the newly discovered O stars outside the confines of Wd 2.  The outlook and our conclusions are summarised in Section 7.

\section{The Data}

We make use of the photometry from two VPHAS+ fields, numbered 1678
and 1679, that are respectively centred on RA 10 18 10.91, Dec -58 03
52.3 (J2000) and on RA 10 25 27.27, Dec -58 03 52.3 (J2000).  These
were observed in succession in the $u$, $g$ and $r$ filters on the
night of 22$^{\rm nd}$ January 2012.  The red filter data in H$\alpha$, $r$ and $i$ were obtained on 29$^{\rm th}$ April 2012.  The seeing, as measured from
the data point spread function, was variable on the earlier night
ranging from 0.62 at best in $g$ up to 1.24 at worst in $r$.  When the
exposures in the red filters were obtained 3 months later, conditions
were more stable, with the typical seeing ranging from 0.8 to 1.0
arcsec. Viewed in comparison to all the VPHAS+ data collected so far,
these observations rank as 2$^{\rm nd}$-quartile quality in $u$ and $g$
(i.e. relatively high quality), and 3$^{\rm rd}$-quartile in $r, i$ and H$\alpha$.
The $5\sigma$ magnitude limits on the single exposures are $u$: 21.0,
$g$: 22.4, $r$: 21.5, $i$: 20.6, and H$\alpha$: 20.4. All magnitudes are in the Vega system. Full details on the survey strategy, the offsets, the exposure times, photometric quality and the data-processing pipeline used are given by \citet{drewetal2014}.

\begin{figure}
\centering
\includegraphics[width=\columnwidth]{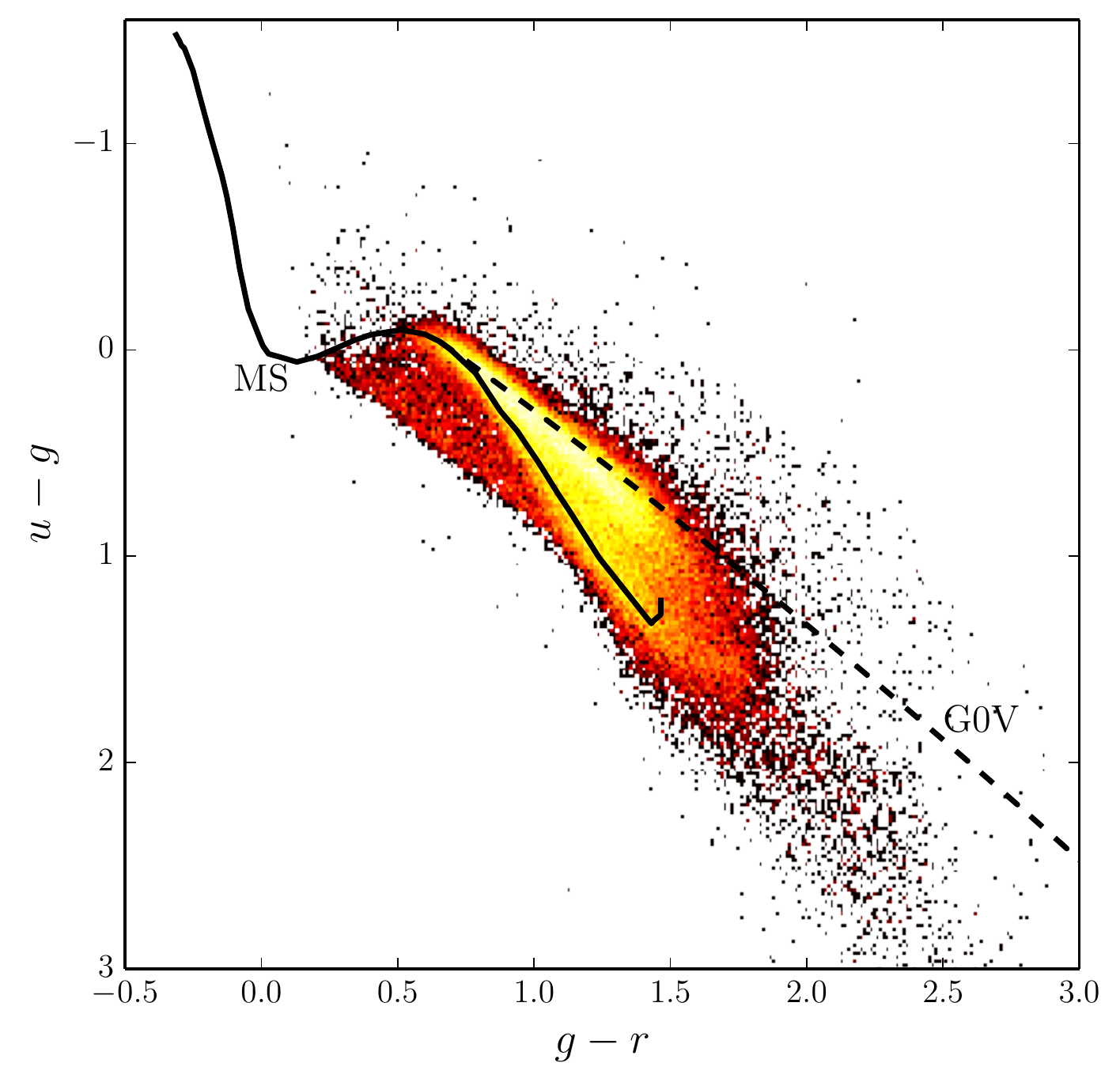}
\caption{Calibration of VPHAS+ data with respect to synthetic reddening tracks from \citet{drewetal2014}. Both the main sequence and G0V reddening vector line up with the main stellar locus.}
\label{fig:stellarlocus}
\end{figure}

Our analysis begins with band-merged catalogues created from the single-band catalogues emerging from the CASU pipeline. In order to correct for the uncertainty in the initial calibration of VPHAS+, a comparison has been made with empirical $g$, $r$ and $i$ observations from the APASS survey and with synthetic tracks in the $(u-g,g-r)$ plane. The median difference between $g$, $r$ and $i$ in the two surveys was applied to the VPHAS+ data. The $u$ band was then calibrated by applying an offset to the $u-g$ scale such that the number density of stars between the synthetic G0V reddening track and the unreddened main sequence is maximised. This ensures that the top and bottom edge of the main stellar locus are aligned with the synthetic tracks as shown in Figure \ref{fig:stellarlocus}. This resulted in offsets relative to the pipeline reduction of $u$: -0.35, $g$: 0.05, $r$: 0.01 and $i$: 0.05 for field 1678 and $u$: -0.34, $g$: 0.06, $r$: 0.01 and $i$: 0.01 for field 1679. With an improved calibration in place, we select stars in the magnitude range $13 < g < 20$ and require random photometric errors to be less than 0.1. Mean magnitudes were taken when repeat photometry was available from the offset fields. Objects were removed if the photometry in the offset field differed by $>0.2$~mags. This removes unreliable photometry due to objects that fall on a CCD edge.

\section{Selection and fitting method}

\subsection{Photometric selection and cross matching}
\label{sec:selection}
We select OB stars using a method that has its origins in the Q Method of \cite{jandm1953}.
On the $(u-g,g-r)$ diagram reddened OB stars of spectral type earlier than B3 are located above and away from the main stellar locus. We initially select our candidate objects above the reddening vector associated with a B3V. In principle no star can be bluer than the Rayleigh-Jeans (RJ) limit which sets an upper bound on the likely location of OB stars in the diagram. The blue objects that lie above the RJ reddening vector were nevertheless included in the selection and their origins are discussed in section \ref{sec:results}.

\begin{figure}
\centering
\includegraphics[width=\columnwidth]{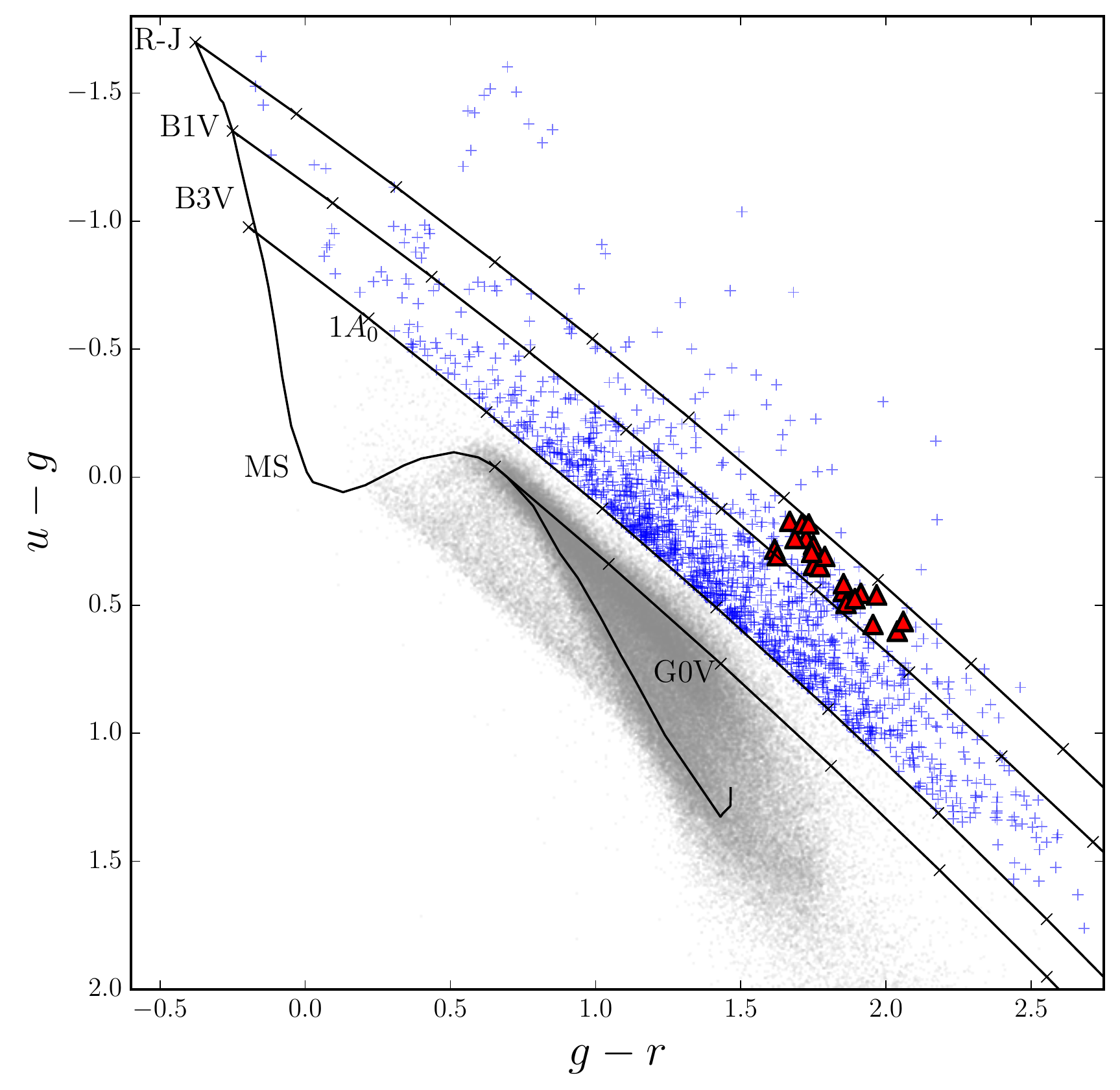}
\caption{Selection of OB stars in and around Wd2. The lower reddening curve is that of a B3V, dropped by 0.1 in u-g in order to capture all early type B stars, and is characterised by an $R_V=3.1$ law. The other reddening curves are that of a B1V and an ideal Rayleigh-Jeans spectrum and are characterised by an $R_V=3.8$ law. Selected OB candidates are blue crosses while the known objects from \protect\cite{Alvarezetal2013} are red triangles.}
\label{fig:OBselect}
\end{figure}

Figure \ref{fig:OBselect} shows the selection of OB candidates (blue crosses) across the two fields as well as the known OB stars from \cite{Alvarezetal2013} that were successfully cross matched with VPHAS+ (shown as red triangles).
Over-plotted are the reddening tracks of a B3V, a B1V and that of a pure RJ spectrum all taken from \cite{drewetal2014}. The tracks we use take into account the measured red leak associated with the $u$-band filter.

Previous results from \cite{Carraroetal2012} and \cite{Alvarezetal2013} suggest an $R_V=3.8$ reddening law is required towards Wd 2. The B1V and RJ reddening vectors have been drawn using this law. To avoid a bias towards this non-standard reddening law we have used the B3V $R_V=3.1$ reddening vector as our lower selection limit and have dropped its position by 0.1\,mags in $u-g$ in order to capture any early B stars that may have been missed. The lower the value of $R_V$, the steeper the reddening vector will be.

Each object was then cross matched to within $1''$ of the best available near infra-red detection in order to access $J, H, K$ photometry. The mean angular cross-match distance was 0.09''. As the stellar density in the central $\sim4'$ of Wd 2 is very high, the \cite{Ascensoetal2007} NIR catalogue was the preferred partner on account of its superior angular resolution. Everywhere else 2MASS was used. This follows the approach taken by \cite{Alvarezetal2013}.

\subsection{SED fitting}

\label{sec:bayesian}

We calculate the probability distribution of a range of model parameters corresponding to a set of empirical measurements, in a Bayesian scheme. This approach is chosen over a straight forward $\chi^2$ minimisation scheme so that we may recover the full posterior probability distribution. This can reveal covariance between different parameters.

Given a set of empirical data, $ d = \lbrace d_1,..,d_i \rbrace $, and a model,  parametrised by a set of parameters, $ \theta = \lbrace \theta_1,.., \theta_i \rbrace $, the \textit{posterior} probability of the parameters can be calculated using Bayes' Theorem:

\begin{equation}
P\left(\theta\mid d \right) = \frac{P\left(d \mid \theta\right)\cdotp P\left(\theta\right)}{P(d)}
\end{equation}

In this expression, $P\left(d \mid \theta\right)$, the \textit{likelihood} is the probability of the data being measured given a set of model parameters. The \textit{posterior} and the \textit{likelihood} are related by the \textit{prior}, $P\left(\theta\right)$, which encodes any known constraints on the model parameters, including known physical bounds. Here $P(d)$ can be treated as a normalising constant and ignored. Hence the \textit{posterior} probability distribution can be found by the relation:

\begin{equation}
P\left(\theta\mid d \right) \propto P\left(d \mid \theta\right)\cdotp P\left(\theta\right)
\end{equation}

In this work the empirical data are derived from the observed SED of each star and they consist of optical and near infrared apparent magnitudes:

\begin{equation}
SED_{obs} = \\ \lbrace u,g,r,i,J,H,K_S\rbrace,
\end{equation}

and their uncertainties:

\begin{equation}
\sigma(SED_{obs}) = \\ \lbrace \sigma_u,\sigma_g,\sigma_r,\sigma_i,\sigma_J,\sigma_H,\sigma_{K_S}\rbrace.
\end{equation}

Along with the random flux errors supplied by the surveys, we have included a systematic uncertainty to account for the independent absolute calibration errors in each band. The values adopted for the latter are 0.04 in the $u$ band, 0.03 in $g, r$ and $i$, 0.03 in the $J$ band and 0.02 in $H$ and $K_{s}$ \citep[see][]{drewetal2014,skrutskieetal2006}.

The model parameters that we are interested in estimating are:

\begin{equation}
\theta = \lbrace \ltf, A_0,R_V, \mu \rbrace
\end{equation}

Where $\ltf$ is the effective temperature, $A_0$ is the monochromatic extinction at 4595$\rm \AA$, $R_V$ is the ratio of total to selective extinction and $\mu$ is the distance modulus.

\subsubsection{Likelihood function}

Defining the likelihood function requires us to define a forward model $SED_{mod}(\theta)$, which predicts the apparent SED of OB stars based on the model parameters $\theta$. The intrinsic SEDs used in the model are taken from the Padova isochrone database \citep[CMD v2.2 \footnote{http://stev.oapd.inaf.it/cgi-bin/cmd};][]{Bressanetal2012,Bertellietal1994} and are supplied in the Vega system. The optical/NIR colours of OB stars do not vary significantly with luminosity class \citep{Martinsetal2005}. Therefore $\log(g)$ was fixed and only main-sequence models were used ($\log(g) \sim 4.0$). Solar metallicity $Z = 0.019$ has been adopted throughout, in view of the fact that the sight lines we explore do not sample a wide range of Galactic radii. This is the same value as used by \cite{Alvarezetal2013}. Fixing these parameters provides a simple grid of absolute magnitude, $M_{\lambda}$, as a function of $\ltf$ in each of the seven bands.

To obtain a continuous grid, each $M_{\lambda}\--\ltf$ relationship was fit with a $2^{nd}$ order polynomial. It can be noted that a linear fit was also trialled but failed to characterize the distributions especially for the low-end values of $\ltf$. Table \ref{table:SEDs} provides sample SEDs.

\begin{table}
\caption{Sample values of the intrinsic SEDs with approximate spectral type equivalents. Magnitudes are in the Vega system. \label{table:SEDs}}

\resizebox{\columnwidth}{!}{%
\begin{tabular}{ccccccccc}
\hline
ST&$\ltf$ & $u$ & $g$ & $r$ & $i$ & $J$ & $H$ & $Ks$ \\
\hline
O3V & 4.65 & -7.32 & -5.78 & -5.48 & -5.33 & -4.88 & -4.73 & -4.63 \\
O9V & 4.50 & -5.28 & -3.86 & -3.60 & -3.45 & -3.03 & -2.90 & -2.80 \\
B1V & 4.40 & -3.97 & -2.70 & -2.47 & -2.34 & -1.98 & -1.85 & -1.77 \\
B3V & 4.27 & -2.31 & -1.33 & -1.16 & -1.07 & -0.80 & -0.70 & -0.65 \\
\hline

\end{tabular}}
\end{table}

The SEDs are then reddened using a \cite{fitzpatrickandmassa2007} reddening law, parametrised by $A_0$ and $R_V$, and then shifted according to a distance modulus. The apparent OnIR SEDs of O and early B stars are largely controlled by these quantities. This is because the OnIR intrinsic colours of OB stars change very slowly as a function of effective temperature \citep{Martinsetal2005}, as the Rayleigh-Jeans limit is approached. This means that $\ltf$ is only weakly constrained, albeit well enough to reach our goal of confirming OB status. As we have no handle on luminosity class, the distance modulus takes the role of a normalisation factor and will also be weakly constrained. In contrast $A_0$ and $R_V$ are very informative and well constrained.

We can now use the forward model to construct a \textit{likelihood} model $P\left(SED_{obs} \mid \theta\right)$ that computes the probability of $SED_{obs}$ given the set of physical parameters $\theta$. Assuming that the uncertainties on the measurements are normally distributed and uncorrelated, this can be described by a multi-variate Gaussian:

\begin{multline}
\label{eqn:matrix}
P \left( SED_{obs} \mid \theta \right) \propto \\ \exp\left[ - \frac{1}{2} \left( SED_{obs}-SED_{mod} \right)^T \Sigma^{-1}\left( SED_{obs}-SED_{mod} \right) \right]
\end{multline}

Where $\Sigma$ is the covariance matrix containing the variance $\sigma^2(SED_{obs})$ in the leading diagonal. In this case Equation \ref{eqn:matrix} reduces to the familiar sum for $\chi^2$:

\begin{equation}
\label{eqn:chi}
P \left( SED_{obs} \mid \theta \right) \propto \exp\left(-\frac{1}{2}\sum\limits_{i}^{n} \frac{(m(obs)_i - m(mod)_i)^2}{\sigma_i^2}\right)
\end{equation}

Where $m(obs)_i$ and $m(mod)_i$ are the observed and model magnitudes in each band $i$.
\subsubsection{Priors}

We adopt a uniform \textit{prior} on each of the model parameters:

\begin{multline}
\label{eqn:params}
P(\theta) = \begin{cases} 1 \quad \text{if} \,\,\begin{cases} 4.2 \,\, \leq \,\, \ltf \,\, \leq \,\, 4.7 \\
0 \,\, \leq \,\, A_0 \,\, \leq \,\, 15 \\
2.1 \,\, \leq \,\, R_V \,\, \leq 5.1 \\
0 \,\, \leq \,\, \mu \,\, \leq \,\, 20
 \end{cases}\\
 0 \quad \text{else}
 \end{cases}
\end{multline}

The upper bound on $\ltf$ is governed by the available models and the lower bound is slightly less than the the typical temperature of a B3V star \citep{zorecandbriot1991} in accordance with our selection in the $\ugr$ diagram. The constraints on $R_V$ are the upper and lower limits measured in the Galaxy \citep{fitzpatrickandmassa2007}. The upper limit on $A_0$ is much larger than maximum extinction allowed for detection of OB stars in VPHAS+ down to $g=20$, assuming a typical rise in visual extinction of 1 magnitude per kpc. This makes the prior on $A_0$ essentially unbound. The upper limit on the distance modulus $\mu$ of 20 is well beyond the realms of the galaxy and so is also essentially unbound. Placing a large but finite limit on $A_0$ and $\mu$ enables the MCMC algorithm to converge more quickly.

\subsubsection{Sampling the posterior distribution using MCMC}

Characterising the \textit{posterior} distribution by computing the probability at all values in the parameter space is computationally expensive. Instead one can sample the distribution using an MCMC algorithm.

In this study we use the Python package \textit{emcee} developed by \cite{Foreman-Mackeyetal2013}. In brief, the software takes a set of parameters and supplies them to a group of \textit{n walkers}. The \textit{walkers} then use a pseudo-random walk to sample the parameter space. At each sample the probability is calculated. By communicating their relative probabilities to one another the \textit{walkers} are able to quickly find and sample the region of high probability without wasting computational time on the parameter combinations of very low probability. The software then returns what are known as \textit{chains} which contain the values of the parameters at every step in the walk. The frequency at which each region in the parameter space is visited is proportional to its probability. The finer details can be found in \cite{Foreman-Mackeyetal2013}.

\section{Validation of Method}

First it is appropriate to verify that our selection method recovers known objects. Second we verify that the fitting algorithm delivers the expected results. To achieve this, we have chosen to compare with the results of the recent study by \cite{Alvarezetal2013}. This is an informative comparison to make both because  this study benefited from the superior angular resolution of HST and because \cite{Alvarezetal2013} used a combination of optical and NIR photometry to derive stellar reddenings as we do here.

\subsection{Photometric selection}
\label{sec:selction_proof}

\begin{figure}
\centering
\includegraphics[width=\columnwidth]{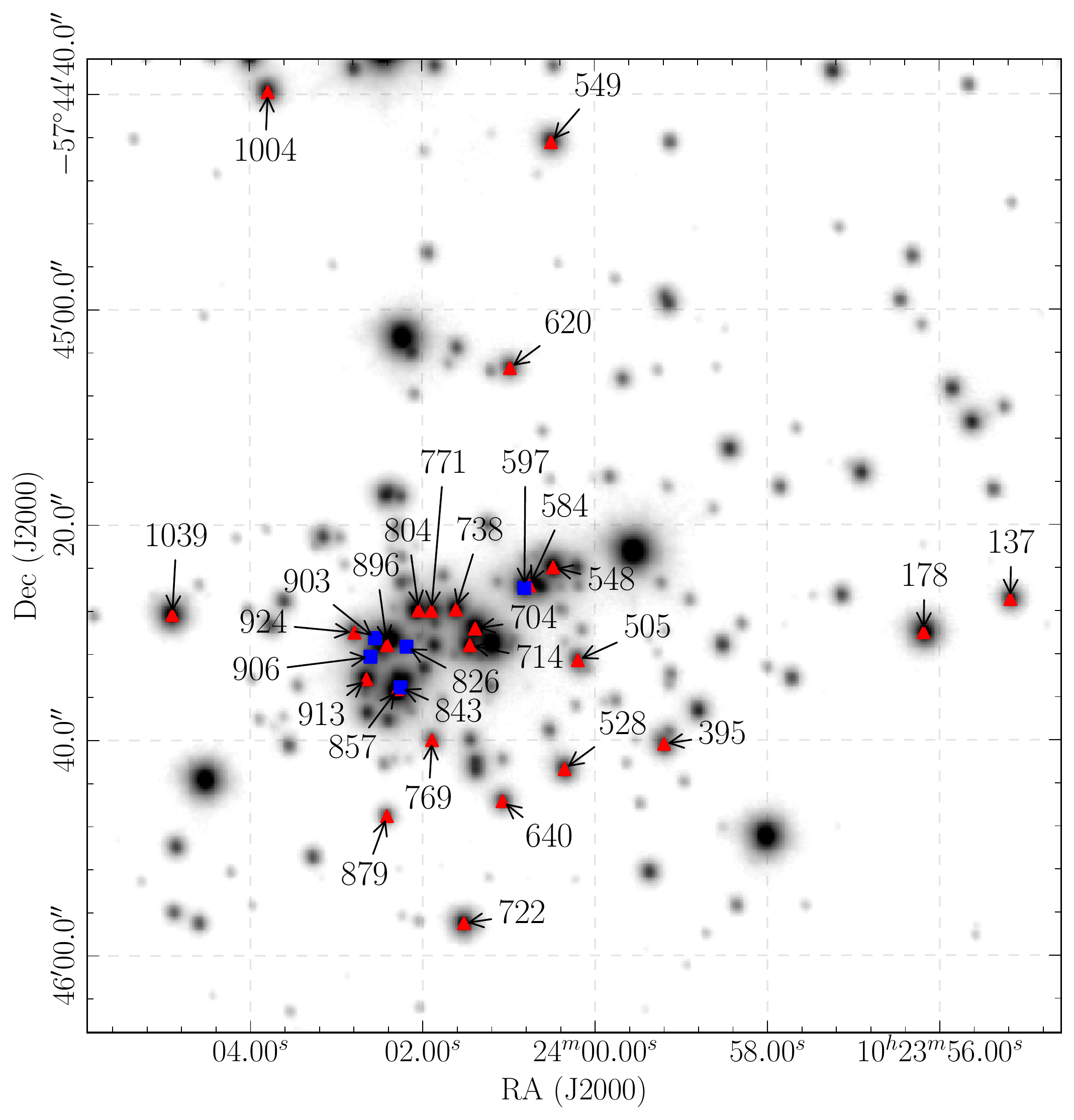}
\caption{Inverse VPHAS+ $g$ band image of the central region of Wd 2 showing the objects with known spectral type from \protect\cite{Alvarezetal2013}. The red triangles are the positions of the objects detected in VPHAS+ and the blue squares are the positions of those that are not detected due to crowding.}
\label{fig:centre}
\centering
\end{figure}

\cite{Alvarezetal2013} derived the extinction properties of 29 known OB stars in the central region of Wd 2, of which, 24 were successfully cross matched with VPHAS+ to within 1". Using the nomenclature from \cite{Alvarezetal2013}, the five missing objects are \#597, \#826, \#843, \#903 and \#906. They appear in some of the most crowded regions of the cluster: the angular resolution of VPHAS+ compared to that of HST is insufficient to separate them from brighter neighbours. Figure \ref{fig:centre} shows the positions of the 24 cross-matched objects and the positions of those that are missing \citep[relative to][]{Alvarezetal2013} over plotted on the g-band image.

\begin{figure}
\centering
\includegraphics[width=\columnwidth]{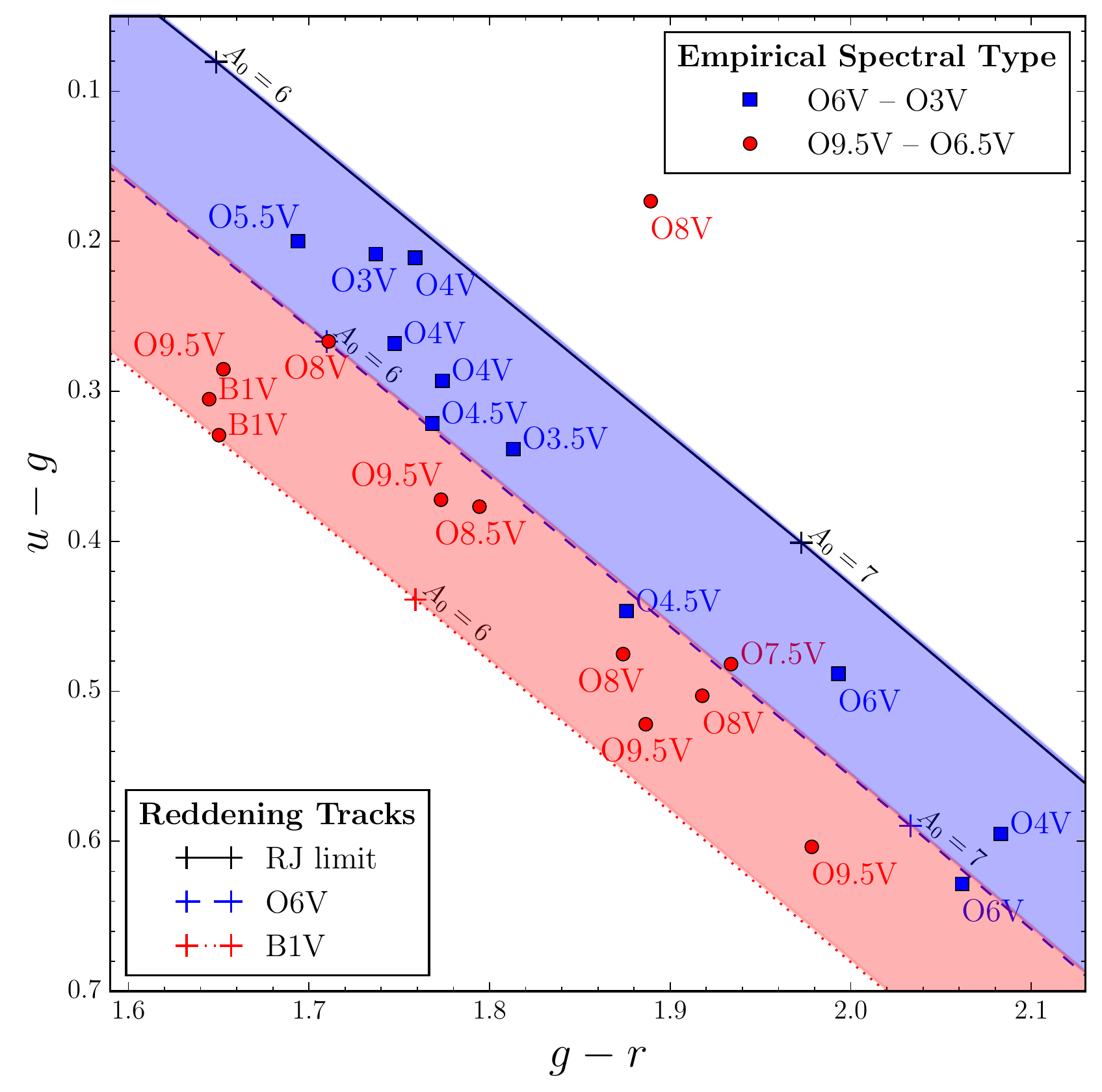}
\caption{Testing the selection process of OB stars associated with Wd2. Objects with known spectral type tend to fall into the correct synthetic spectral type range with an $R_V=3.8$ reddening law.}
\label{fig:knownst}
\centering
\end{figure}

Figure \ref{fig:knownst} is the highly magnified section of Figure \ref{fig:OBselect} that contains the objects with known spectral type. The red and blue shaded regions are where we expect to find late-type (O9 - O6) and early-type (O6 - RJ) O stars respectively. We find that the majority of the objects are correctly separated into their respective early or late spectral-type zones defined by the $R_V=3.8$ reddening tracks. This gives an early indication that an $R_V \sim 3.8$ reddening law is required for this sight-line and that the calibration of the data is in good agreement with the synthetic photometry. 

Object \#771 falls well above the `RJ limit'.  As a confirmed O8V star, its position in the $\ugr$ diagram is clearly anomalous. Close inspection of the image suggests that the photometry of this star is affected by a bright neighbour.

\subsection{SED fitting}
\label{sec:sed_proof}

Ultimately 21 of the 24 known objects were suitable for SED fitting. These objects are tabulated in Table \ref{table:compare}. Two of the objects left out are \#896 and \#771 for which there is no detection in one or more of the the optical bands due to blending. The third is object \#1004 for which the near-infrared photometry is incomplete.

\begin{figure}
\begin{center}
\includegraphics[width=\columnwidth]{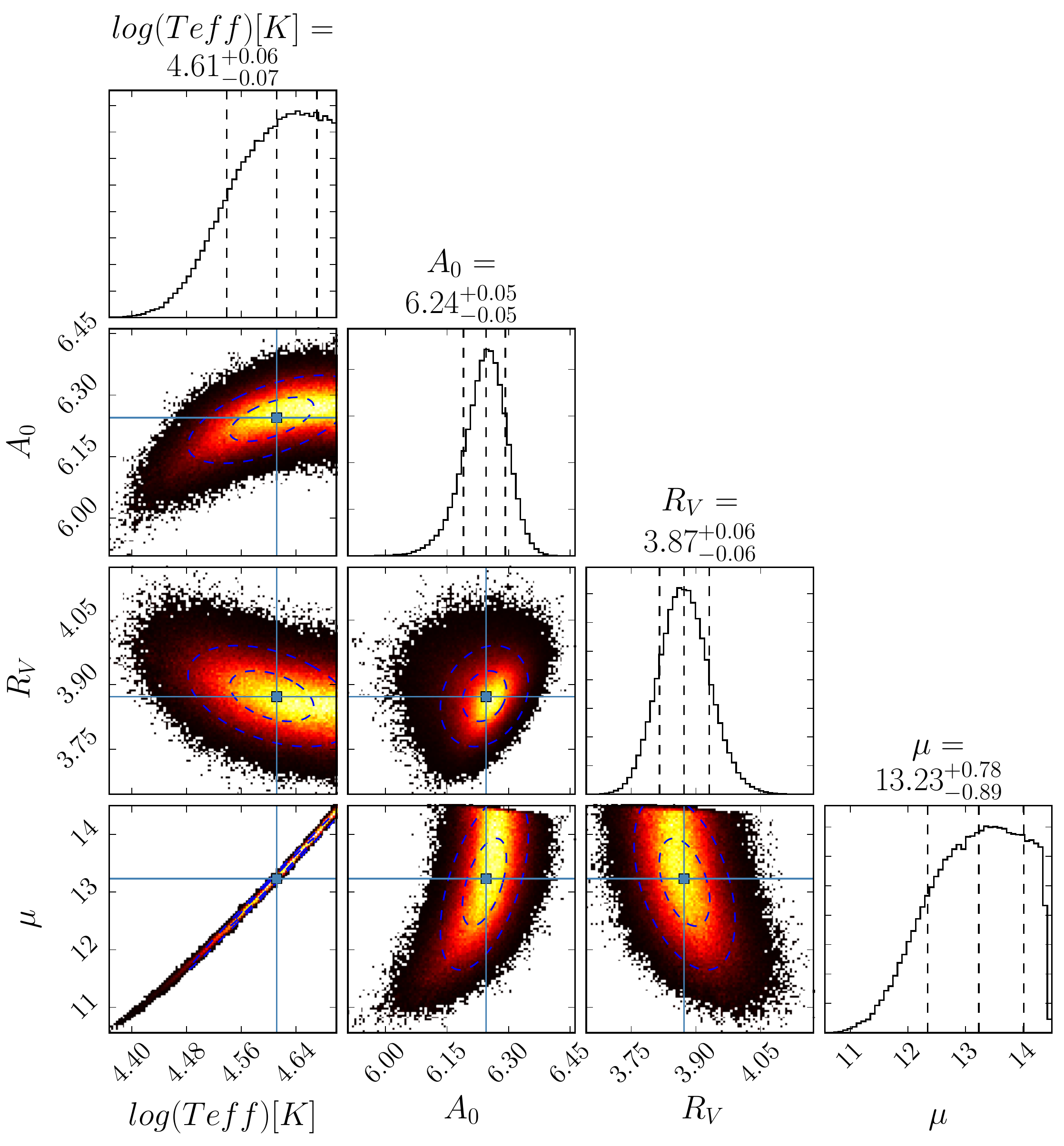}
\includegraphics[width=\columnwidth]{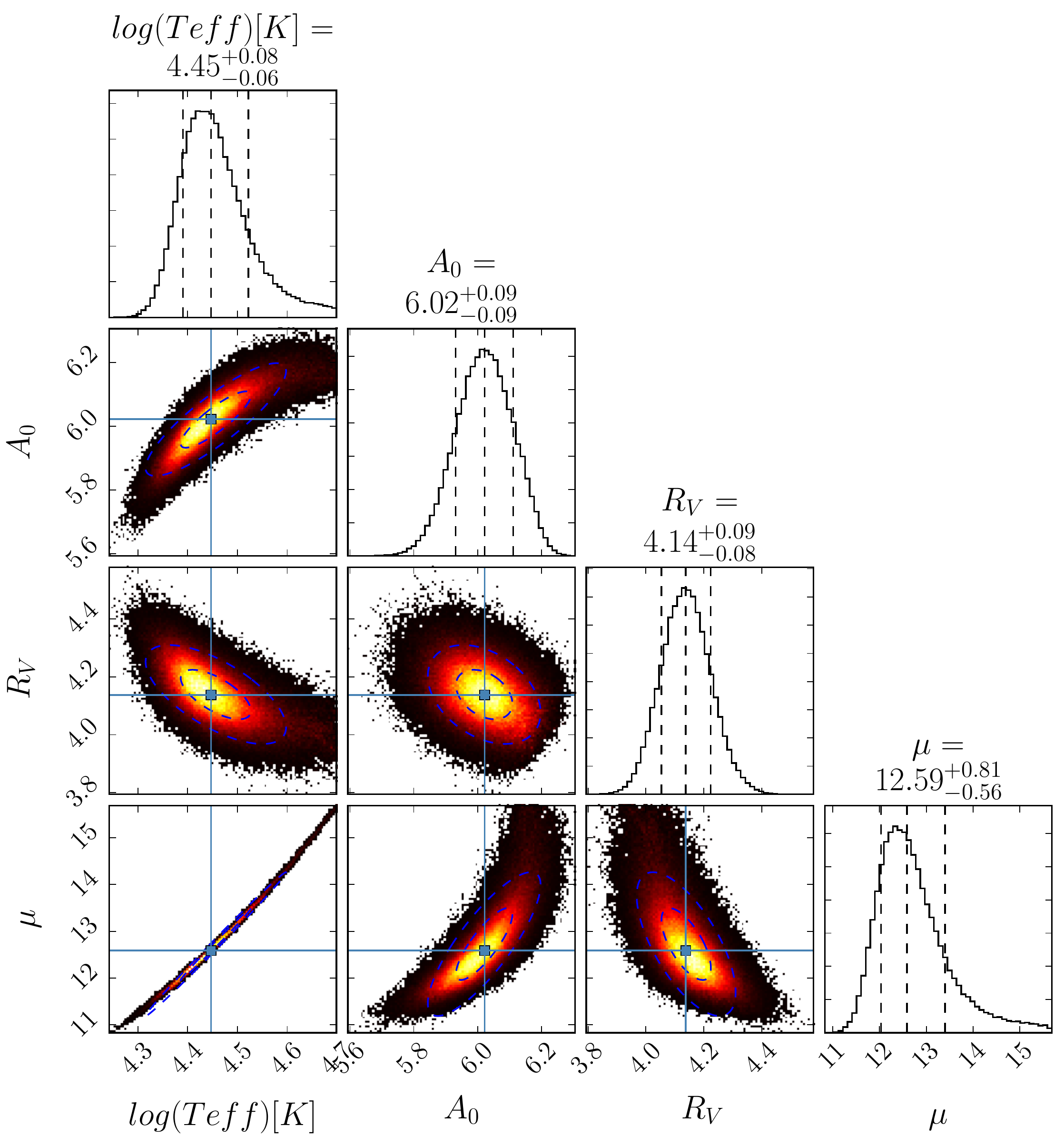}

\caption{PDFs of the fitting parameters as a result of the MCMC simulation for stars \#913 an O4V (top) and \#549 and B1V (bottom) using the numbering system from \protect\cite{Alvarezetal2013}.}

\label{fig:MCMC}
\end{center}
\end{figure}

For each of the 21 objects for which we have computed SED fits, the posterior distribution was sampled with 100 \textit{walkers} over 10000 \textit{iterations} with a 1000 iteration burn in. The typical autocorrelation time for each walk (or number of steps per independent sample) was found to be well below 100, which indicates that the posteriors are thoroughly sampled. We can determine the probability distributions for each parameter by marginalising over all other parameters.
We visualize this by constructing 1-D histograms of the values of each parameter visited in the random walk.
We can also check for covariance or degeneracy between parameters by constructing marginalised 2-D histograms for each pair of parameters. Figure \ref{fig:MCMC} shows an example of these diagrams for an O4V and a B1V star in the sample (\#913 and \#549).

The obvious difference between the two cases is apparent in the 1-D marginalisation of parameters. We see that the hotter the object the more skewed the probability distributions in $\ltf$ and $\mu$ become. This can be attributed to the fact that the hotter SEDs are approaching the RJ tail. This makes it more difficult to differentiate the temperature of the hottest stars and consequently the luminosity and distance. This makes the drop off in probability at the hot end more shallow. This intrinsic feature also means that the uncertainties on $\ltf$ and $\mu$ increase with temperature but has the positive effect of decreasing the uncertainties on $A_0$ and $R_V$. For the later type stars $\ltf$ is better defined but still uncertain.

The value adopted for each parameter is the median of the marginalised posterior distribution with upper and lower uncertainties defined by the $16^{\rm th}$ and $84^{\rm th}$ percentiles. We find that we are able to determine the values of $A_0$ and $R_V$ with relatively high precision (better than $\pm 0.09$\,mag and $\pm 0.08$ respectively in all cases). These uncertainties are similar to those found by \cite{Alvarezetal2013}. We note that $R_V$ and $A_0$ are well defined and show negligible covariance relative to each other and only modest covariance with respect to $\ltf$ and $\mu$.

However, as expected, our determination of temperature and distance are not so informative. For object \#913, $\ltf=4.61^{+0.06}_{-0.07}$ and $\mu=13.23^{+0.78}_{-0.89}$. This corresponds to values of $T_{\rm eff}=40.7^{+6.0}_{-6.1}$kK, or a spectral type range from O8V to O2V. The results for $\mu$ translate to $d=4.4^{+1.9}_{-1.5}$kpc. This already significant distance uncertainty is nevertheless an underestimate given that neither the luminosity class or metallicity uncertainties have been formally incorporated. In addition we are treating all stars as if single which biases the inferred distance moduli to lower values by up to 0.75 mag. Because of the relative lack of constraint on $\ltf$ from the intrinsic colours of OB stars, the error in $\ltf$ is driven mainly by the error in $\mu$. In comparison the direct effect of binarity on $\ltf$, through colour-changes, will be small. It is plainly apparent in Figure \ref{fig:MCMC} that $\ltf$ and $\mu$ are strongly and positively covariant. The role of the distance modulus is essentially that of a normalisation parameter.

\begin{figure}
\begin{center}
\includegraphics[width=\columnwidth]{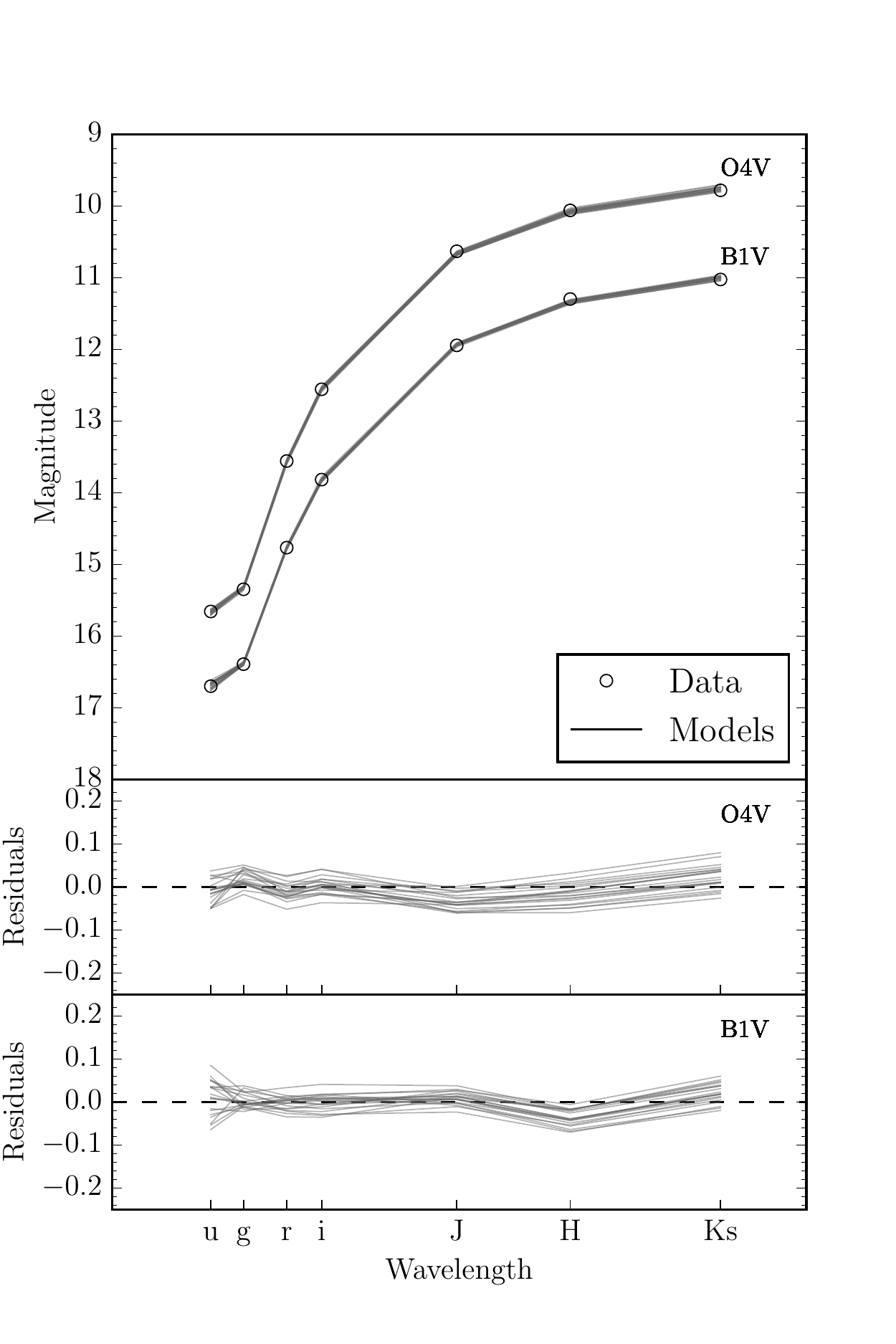}
\caption{Visualisation of the posterior distributions
 of objects \#913 and \#549 \protect\citep[from][]{Alvarezetal2013} in SED data space. The top panel shows 30 model SEDs for both objects (gray solid lines),
 generated from a random sampling of the posterior parameter distributions
 shown in Figure \ref{fig:MCMC}.
 Our photometric data is plotted on top (circles).
 The bottom panels show the residuals.}
\label{fig:sed}
\end{center}
\end{figure}

Figure \ref{fig:sed} shows the results for the O4V star from Figure \ref{fig:MCMC} translated into the original SED data space.
The top panel shows the observed SED over-plotted by 30 randomly sampled model SEDs that are drawn from the posterior distributions shown in Figure \ref{fig:MCMC}. The lower panel shows the residuals between them.
We can see that for each band, across all the posterior distributions, the differences between the models and the data never exceed $\sim0.1$\,mag. The discrepancies between the model and data can be attributed to one or more of the following: inaccuracies in the intrinsic SEDs of OB stars in the Padova isochrones; inaccuracies in the shape of the reddening law; a calibration offset between the optical and NIR catalogues.

\begin{table*}

\captionof{table}{Table comparing the derived stellar parameters of objects with known spectral type from \protect\cite{Alvarezetal2013} with the results in this study. The ID given corresponds to the numeration given by \protect\cite{Alvarezetal2013} \label{table:compare}. Most of the effective temperatures in the HST column were derived spectroscopically by \protect\cite{Alvarezetal2013} and uncertainties were given. The rest have no provided uncertainty as they were estimated from their spectral types using the temperature scales from \protect\cite{Martinsetal2005} and \protect\cite{zorecandbriot1991}.}

\resizebox{\textwidth}{!}{%
\begin{tabular}{cccccccccccc}

\hline
\bf ID &\bf ST &\multicolumn{2}{c}{$\bf A_V$} & \multicolumn{2}{c}{$\bf R_V$} &
\multicolumn{2}{c}{$\bf log(T_{eff})$} &
\multicolumn{2}{c}{$\bf \mu$} &
\multicolumn{2}{c}{$\bf V$}\\
& & VPHAS+ & HST & VPHAS+ & HST & VPHAS+ & HST & VPHAS+ & HST & VPHAS+ & HST \\
\hline\noalign{\smallskip}
137 & O4  V & $7.47^{+0.04}_{-0.04}$ & $7.41\pm0.22$ & $4.05^{+0.05}_{-0.05}$ & $3.84\pm0.07$ & $4.63^{+0.05}_{-0.06}$ & $4.633\pm0.004$ & $13.43^{+0.62}_{-0.79}$ & $13.19\pm0.45$ & $15.496\pm+0.056$ & $15.591\pm0.006$ \\ [0.2cm]
178 & O4  V-III((f)) & $6.34^{+0.04}_{-0.04}$ & $6.38\pm0.07$ & $4.03^{+0.06}_{-0.06}$ & $3.93\pm0.03$ & $4.63^{+0.05}_{-0.07}$ & $4.629\pm0.002$ & $13.38^{+0.64}_{-0.82}$ & $11.79\pm0.16$ & $14.385\pm+0.055$ & $14.490\pm0.004$ \\[0.2cm]
395 & O7.5V & $6.78^{+0.07}_{-0.08}$ & $6.92\pm0.07$ & $4.08^{+0.07}_{-0.07}$ & $3.77\pm0.03$ & $4.52^{+0.09}_{-0.07}$ & $4.544\pm0.000$ & $12.91^{+1.09}_{-0.76}$ & $12.78\pm0.18$ & $15.688\pm+0.056$ & $16.019\pm0.062$ \\[0.2cm]
505 & O8.5V & $6.19^{+0.05}_{-0.06}$ & $6.36\pm0.14$ & $3.84^{+0.06}_{-0.06}$ & $3.71\pm0.06$ & $4.59^{+0.07}_{-0.08}$ & $4.531\pm0.006$ & $14.57^{+0.93}_{-0.95}$ & $13.29\pm0.30$ & $15.889\pm+0.056$ & $16.094\pm0.005$ \\[0.2cm]
528 & O8  V & $6.72^{+0.06}_{-0.07}$ & $6.97\pm0.14$ & $4.02^{+0.07}_{-0.06}$ & $3.99\pm0.05$ & $4.56^{+0.09}_{-0.08}$ & $4.544\pm0.005$ & $13.34^{+1.10}_{-0.89}$ & $12.55\pm0.30$ & $15.571\pm+0.056$ & $15.841\pm0.005$ \\[0.2cm]
548 & O4  V & $6.34^{+0.05}_{-0.05}$ & $6.48\pm0.10$ & $4.02^{+0.06}_{-0.06}$ & $3.76\pm0.04$ & $4.61^{+0.06}_{-0.07}$ & $4.633\pm0.002$ & $13.11^{+0.81}_{-0.90}$ & $13.19\pm0.23$ & $14.361\pm+0.055$ & $14.522\pm0.002$ \\[0.2cm]
549 & B1  V & $6.02^{+0.09}_{-0.09}$ & $6.09\pm0.08$ & $4.14^{+0.09}_{-0.08}$ & $4.01\pm0.04$ & $4.45^{+0.08}_{-0.06}$ & $4.398\pm0.000$ & $12.59^{+0.81}_{-0.56}$ & $11.68\pm0.19$ & $15.485\pm+0.056$ & $15.562\pm0.005$ \\[0.2cm]
584 & O8  V & $4.60^{+0.04}_{-0.04}$ & $6.19\pm0.05$ & $2.91^{+0.04}_{-0.04}$ & $3.73\pm0.02$ & $4.66^{+0.03}_{-0.05}$ & $4.544\pm0.002$ & $15.24^{+0.42}_{-0.64}$ & $12.94\pm0.12$ & $14.195\pm+0.055$ & $15.442\pm0.004$ \\[0.2cm]
620 & B1  V & $5.77^{+0.09}_{-0.09}$ & $5.77\pm0.08$ & $4.00^{+0.08}_{-0.08}$ & $3.82\pm0.04$ & $4.46^{+0.08}_{-0.06}$ & $4.398\pm0.000$ & $13.46^{+0.86}_{-0.56}$ & $12.56\pm0.19$ & $16.007\pm+0.057$ & $16.086\pm0.006$ \\[0.2cm]
640 & O9.5V & $6.32^{+0.05}_{-0.07}$ & $6.37\pm0.05$ & $3.97^{+0.07}_{-0.06}$ & $3.73\pm0.02$ & $4.57^{+0.08}_{-0.08}$ & $4.505\pm0.002$ & $14.30^{+1.07}_{-0.91}$ & $13.11\pm0.13$ & $16.065\pm+0.057$ & $16.234\pm0.006$ \\[0.2cm]
704 & O4  V & $6.03^{+0.05}_{-0.05}$ & $6.27\pm0.29$ & $3.94^{+0.06}_{-0.06}$ & $3.76\pm0.12$ & $4.61^{+0.06}_{-0.07}$ & $4.681\pm0.008$ & $12.91^{+0.77}_{-0.85}$ & $14.26\pm0.63$ & $13.844\pm+0.055$ & $14.059\pm0.002$ \\[0.2cm]
714 & O3  V & $5.61^{+0.04}_{-0.05}$ & $6.08\pm0.11$ & $3.67^{+0.06}_{-0.06}$ & $3.73\pm0.05$ & $4.62^{+0.05}_{-0.07}$ & $4.643\pm0.000$ & $14.29^{+0.69}_{-0.82}$ & $14.53\pm0.26$ & $14.642\pm+0.055$ & $15.017\pm0.003$ \\[0.2cm]
722 & O6  V & $7.21^{+0.05}_{-0.06}$ & $7.23\pm0.04$ & $3.94^{+0.06}_{-0.05}$ & $3.65\pm0.01$ & $4.61^{+0.06}_{-0.08}$ & $4.584\pm0.001$ & $12.79^{+0.84}_{-0.94}$ & $12.04\pm0.11$ & $14.944\pm+0.055$ & $15.060\pm0.030$ \\[0.2cm]
738 & O5.5V & $5.84^{+0.05}_{-0.05}$ & $6.02\pm0.08$ & $3.88^{+0.06}_{-0.06}$ & $3.73\pm0.04$ & $4.61^{+0.06}_{-0.07}$ & $4.602\pm0.000$ & $13.91^{+0.81}_{-0.90}$ & $13.39\pm0.19$ & $14.696\pm+0.055$ & $14.896\pm0.003$ \\[0.2cm]
769 & O9.5V & $6.50^{+0.06}_{-0.09}$ & $6.63\pm0.06$ & $3.86^{+0.07}_{-0.06}$ & $3.65\pm0.02$ & $4.54^{+0.10}_{-0.08}$ & $4.491\pm0.002$ & $14.12^{+1.20}_{-0.89}$ & $13.04\pm0.13$ & $16.351\pm+0.057$ & $16.576\pm0.008$ \\[0.2cm]
804 & O6  III & $7.11^{+0.11}_{-0.11}$ & $6.91\pm0.04$ & $4.11^{+0.09}_{-0.09}$ & $3.71\pm0.01$ & $4.60^{+0.07}_{-0.08}$ & $4.582\pm0.001$ & $12.14^{+0.89}_{-0.96}$ & $11.78\pm0.10$ & $14.290\pm+0.055$ & $14.433\pm0.003$ \\[0.2cm]
857 & O4.5V & $6.50^{+0.08}_{-0.08}$ & $6.13\pm0.08$ & $4.17^{+0.09}_{-0.08}$ & $3.63\pm0.03$ & $4.56^{+0.09}_{-0.08}$ & $4.623\pm0.002$ & $11.30^{+1.08}_{-0.91}$ & $12.65\pm0.18$ & $13.335\pm+0.055$ & $13.869\pm0.003$ \\[0.2cm]
879 & O9.5V & $6.77^{+0.06}_{-0.07}$ & $6.98\pm0.07$ & $3.82^{+0.06}_{-0.06}$ & $3.70\pm0.03$ & $4.57^{+0.08}_{-0.08}$ & $4.519\pm0.003$ & $14.37^{+1.05}_{-0.98}$ & $13.11\pm0.16$ & $16.510\pm+0.058$ & $16.645\pm0.056$ \\[0.2cm]
913 & O3-4V & $6.23^{+0.05}_{-0.06}$ & $6.42\pm0.11$ & $3.87^{+0.06}_{-0.06}$ & $3.66\pm0.04$ & $4.61^{+0.06}_{-0.07}$ & $4.642\pm0.002$ & $13.23^{+0.78}_{-0.89}$ & $13.45\pm0.24$ & $14.344\pm+0.055$ & $14.531\pm0.002$ \\[0.2cm]
924 & O8  V & $6.25^{+0.06}_{-0.07}$ & $6.40\pm0.07$ & $3.68^{+0.06}_{-0.06}$ & $3.60\pm0.03$ & $4.57^{+0.08}_{-0.08}$ & $4.544\pm0.000$ & $14.08^{+1.02}_{-0.91}$ & $13.16\pm0.16$ & $15.680\pm+0.056$ & $15.960\pm0.005$ \\[0.2cm]
1039 & O4-5V & $6.43^{+0.05}_{-0.05}$ & $6.42\pm0.10$ & $3.80^{+0.06}_{-0.05}$ & $3.47\pm0.04$ & $4.62^{+0.05}_{-0.07}$ & $4.622\pm0.002$ & $13.26^{+0.70}_{-0.86}$ & $12.98\pm0.22$ & $14.429\pm+0.055$ & $14.523\pm0.030$ \\[0.2cm]
\hline
\end{tabular}}

\end{table*}

\begin{figure}
\begin{center}
\includegraphics[width=\columnwidth]{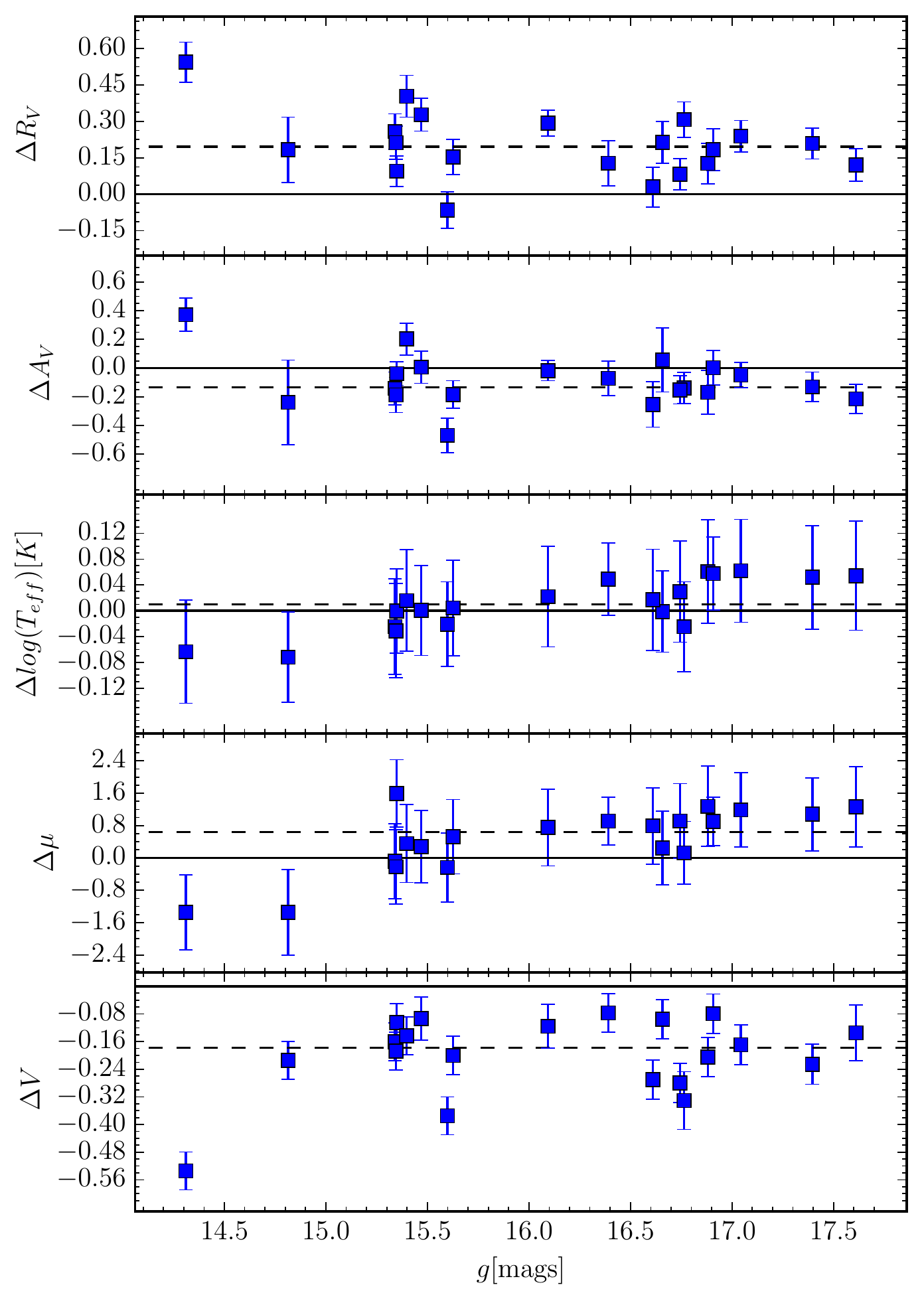}
\caption{The difference between stellar parameters found in this study and those found by \protect\cite{Alvarezetal2013}. The solid line shows zero difference while the dashed line shows the median difference.}
\label{fig:hstcompare}
\end{center}
\end{figure}

Table \ref{table:compare} compares the stellar parameters of the 21 known OB stars derived in this study with those from \cite{Alvarezetal2013}. Here $A_0$ has been converted to $A_V$ and the VPHAS+ $g$ band magnitudes have been converted to $V$ band using the Sloan to Johnson conversion from Lupton (2005)\footnote{https://www.sdss3.org/dr8/algorithms/sdssUBVRITransform.php} for ease of comparison. We also note that our SED-derived $\ltf$ values are compared to spectroscopic values where available \citep{Alvarezetal2013,Rauwetal2007}. Otherwise effective temperatures are derived from spectral types according to the temperature scales of \cite{Martinsetal2005} and \cite{zorecandbriot1991}. We restrict our comparison to the results in \cite{Alvarezetal2013} based on the \cite{fitzpatrickandmassa2007} extinction curves.

Figure \ref{fig:hstcompare} plots the difference between the values derived in the two studies. It must be noted that star \#584 has not been included in this analysis as extreme blending has substantially affected its photometry (see Figure \ref{fig:centre} and Table \ref{table:compare}).

A significant difference is found between the transformed $V$ band magnitudes in VPHAS+ and HST of $\sim0.18$\,mag, such that VPHAS+ is brighter. \cite{Alvarezetal2013} compare their empirical $B$ and $V$ band measurements with those of \cite{Moffatetal1991} and \cite{Rauwetal2007} and find that those ground based measurements are also systematically brighter, by $0.18$ and $0.15$\,mag, and by $0.22$ and $0.12$ respectively. \cite{Alvarezetal2013} suggest that the difference may be due to source blending following on from the effects of atmospheric seeing. If this were the case we would expect to find objects in the most crowded/blended region of the cluster to be consistently more discrepant. As we do not see this effect we suspect a real calibration difference. \cite{huretal2015} have also uncovered a similar problem but find good agreement between their optical photometry and that of \cite{Rauwetal2007}. If the scale of \cite{Rauwetal2007} is the right one, our photometry may be too bright by $\sim0.05$ mag.

The apparent systematic calibration difference between the two data sets is reflected in the derived values of $A_V$. In particular the median of the star-by-star differences in $A_V$ shows that our extinctions are on average $0.14$\,mag less than those derived by \cite{Alvarezetal2013}. The median $A_V$ with the $16^{\rm th}$ and $84^{\rm th}$ percentiles in this study and in \cite{Alvarezetal2013} are; $A_V(\rm VPHAS+)=6.34^{-0.32}_{+0.44}$ and $A_V(\rm HST)=6.41^{-0.32}_{+0.56}$. With brighter optical magnitudes there is also an offset in $R_V$ such that our values are higher: the median star-by-star difference in $R_V$ is 0.20, while sample medians are respectively $R_V(\rm VPHAS+)=3.96^{-0.14}_{+0.12}$ and $R_V(\rm HST)=3.73^{-0.08}_{+0.11}$.

Despite the expectation of poor constraints on distance, the difference in the median values of $\mu$ happen to be very small: $\rm \mu(VPHAS+)=13.36^{-0.57}_{+0.92}$ and $\rm \mu(HST)=13.07^{-1.02}_{+0.31}$. This is likely to be due to the O stars in Wd 2 being on the main sequence, matching our assumption. Similarly there is only a modest offset on average in the measures of effective temperature.

The results of this comparison are encouraging. We have found good quantitative agreement, within the uncertainties, between our derived parameters and those of \cite{Alvarezetal2013} drawing on HST optical photometry. Where there are differences, we understand their origin. This gives us confidence that both our method and the underlying VPHAS+ data are producing reliable results.

\section{Results}
\label{sec:results}
Here we apply the SED fitting methods discussed above to the full selection of OB candidates from our pilot $\sim$2 sq.deg field.

\subsection{`Goodness-of-fit'}

The posterior distributions obtained tell us the most probable parameters given the data, however they do not tell us anything about `goodness-of-fit'. As some objects in our selection may be contaminants or may just have bad photometry, it is important to determine how well the data fit the model in order to obtain a `clean' selection of OB stars. We have opted to use the value of $\chi^2$, given by the SED fits, at the median values in the marginalised posterior distribution. We are aware that the posterior medians may not exactly trace the maximum likelihood, but they provide a representative sample.

\begin{figure}
\begin{center}
\includegraphics[width=\columnwidth]{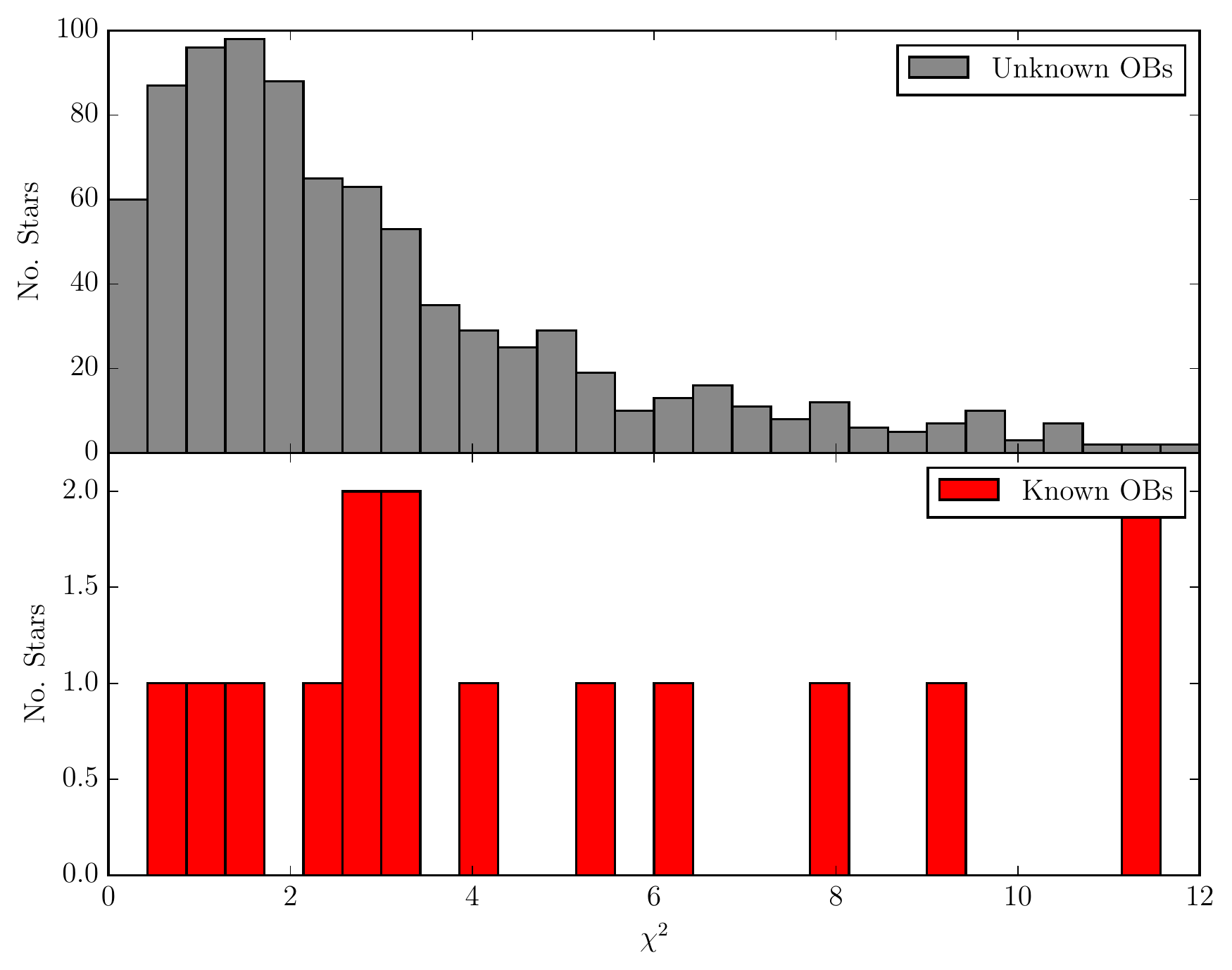}
\caption{$\chi^2$ distributions for the known objects (bottom) and the wider selection (top). The $\chi^2$ distribution for the wider selection peaks at $\sim 1$ as a expected from a distribution with $k=3$ degrees of freedom. Using a $5\%$ significance level we judge objects with $\chi^2>7.82$ to be unsatisfactorily fit. The known objects with poor fits are subject to photometric blending in the cluster's core.}
\label{fig:chi_hist}
\end{center}
\end{figure}

Figure \ref{fig:chi_hist} shows the $\chi^2$ distribution of the fits to all
1050 objects in the wider selection above the distribution obtained for the known objects from \cite{Alvarezetal2013}. Since we are fitting 7 data points with 4
parameters we expect a $k = 3$ $\chi^2$ distribution peaking at 1 --
the top panel of Figure \ref{fig:chi_hist} indicates this is what
happens and, by implication, that the uncertainties on our data points
are not significantly over- or under- estimated. In keeping with this,
we have chosen to use the commonly adopted $5\%$ significance level,
at $\chi^2 = 7.82$ as the limit beyond which we judge the fits to the
applied model to be unsatisfactory.  This cut makes reasonable sense
when applied to the $\chi^2$ distribution for the known objects
\citep[in common with][]{Alvarezetal2013}, in that the 10 confirmed
OB stars beyond the chosen cut are mainly there because of the impact
on the photometry of the blending in the crowded central parts of Wd 2
present in the VST data. For this reason we have still tabulated those objects not meeting our selection criteria but have not used them in any further analysis. We note that if both 2MASS \cite{skrutskieetal2006} and \cite{Ascensoetal2007} photometry are available we keep which ever yields a better $\chi^2$.

\subsubsection{Further cross-matches with previously catalogued objects \label{sec:simbad}}

All of the objects in the initial selection were cross-matched to
$<1''$ with the SIMBAD \citep{wengeretal2000} database to check for further examples of
objects of known type.

\cite{Tsujimotoetal2007} conducted a $17 \times
17$ arcmin high resolution X-ray imaging survey centred on Wd2 and the
surrounding star forming region RCW 49.  They identified 17 new X-ray
emitting OB candidates in this larger region, enclosing that studied
by \cite{Alvarezetal2013}. On using a 1" cross match radius we find 8
of these objects make it into our selection. Five of the missing
objects are picked up by VPHAS+ but have $g < 13$ and hence were too bright to be selected.  Conversely, the remaining 4 objects are detected by VPHAS+ but are too faint ($g>20$) to be in our selection. It is likely that these objects are highly reddened.

Across all other literature sources, accessed via SIMBAD, fourteen further stars of
confirmed type were found (see Table \ref{table:simbad}). The breakdown of their classifications is
as follows: six stars with a Wolf-Rayet (WR) component, three OV, two OIII, one OVb, one B5Vne, one carbon star and one star listed as M1III. All six WR stars, the carbon star and one of the OV stars could not be
fitted convincingly as reddened OB stars (i.e. $\chi^2>7.82$), while
the others were ($\chi^2<7.82$). The OVb was confirmed as an O3V + O5.5V binary system by \cite{Alvarezetal2013} but was not used in their SED fitting analysis -- hence it did not feature in Section \ref{sec:sed_proof}. On close inspection of the literature, it became clear that the SIMBAD M1III attribution matching one of our selected objects is wrong, resulting from confusion over the sky
position of the previously catalogued HAeBe candidate, THA 35-II-41.
THA 35-II-41 is indeed one of our selected objects but it is not at
the position attributed to it by \cite{Carmonaetal2010} where these
authors observed an M giant spectrum.

We also detect seven bright objects in the originally NIR selected open cluster DBS2003 45 \citep{dbs2003} centred at 10h19m10.5s $-58\degree 02'22. 6''$. The study by \cite{Zhuetal2009} identifies seven OB stars in this cluster estimated as ranging from spectral type B0 to O7 from low resolution NIR spectroscopy. However, six out of seven of the positions given in Table 2 of \cite{Zhuetal2009} do not match with the VPHAS+ positions nor with any detections in the 2MASS point source catalogue. We therefore suspect that there is an error in the positions that they give whilst our objects are in common. We find these are among the most highly extinguished objects in our selection with an average $A_V=8.37$.

\begin{table*}
\caption{Objects crossed matched with SIMBAD in the selection which have known spectral type. Derived parameters of highly evolved objects will be inaccurate due to the main-sequence assumption as shown by their large $\chi^2$ values. On further inspection of the literature the classification object \#895 is much different from that in SIMBAD(see Section \label{table:simbad})}

\resizebox{\textwidth}{!}{%
\begin{tabular}{cccccccccccc}

\hline
ID&RA&DEC&Identifier&Spectral Type&$g$&$\log(T_{eff})$&$R_V$&$A_0$&$\mu$ &$\chi^2$ \\

\hline\noalign{\smallskip}

282&10 18 04.98&-58 16 26.27&WR  19                            &WC5+O9             &$14.02$&$4.37^{+0.05}_{-0.04}$&$5.79^{+0.09}_{-0.09}$&$4.14^{+0.09}_{-0.08}$&$9.60^{+0.48}_{-0.38}$&$39.60$\\ \noalign{\smallskip}
335&10 18 53.39&-58 07 52.94&WR  19a                           &WN                 &$15.45$&$4.54^{+0.10}_{-0.08}$&$8.59^{+0.06}_{-0.09}$&$4.32^{+0.07}_{-0.06}$&$9.87^{+1.19}_{-0.89}$&$10.21$\\ \noalign{\smallskip}
437&10 20 17.50&-57 44 59.39&C* 1665                           &C*                 &$16.54$&$4.66^{+0.02}_{-0.03}$&$12.30^{+0.04}_{-0.04}$&$4.54^{+0.03}_{-0.03}$&$8.05^{+0.32}_{-0.39}$&$436.47$\\ \noalign{\smallskip}
560&10 22 05.75&-57 53 46.03&2MASS J10220574-5753460           &B5Vne              &$15.71$&$4.42^{+0.07}_{-0.05}$&$5.64^{+0.09}_{-0.09}$&$3.80^{+0.07}_{-0.07}$&$11.98^{+0.68}_{-0.49}$&$1.53$\\ \noalign{\smallskip}
644&10 23 23.50&-58 00 20.80&SS 215                            &O2If*/WN5          &$13.48$&$4.41^{+0.06}_{-0.05}$&$5.67^{+0.09}_{-0.09}$&$4.27^{+0.10}_{-0.09}$&$9.60^{+0.57}_{-0.44}$&$16.06$\\ \noalign{\smallskip}
687&10 23 58.01&-57 45 48.93&V* V712 Car                       &O3If*/WN6+O3If*/WN6&$14.48$&$4.48^{+0.08}_{-0.06}$&$7.50^{+0.08}_{-0.09}$&$4.27^{+0.07}_{-0.07}$&$9.39^{+0.91}_{-0.63}$&$10.47$\\ \noalign{\smallskip}
717&10 24 01.20&-57 45 31.03&Cl* Westerlund    2    MSP     188&O3V+O5.5V          &$14.34$&$4.53^{+0.10}_{-0.08}$&$6.79^{+0.09}_{-0.09}$&$4.41^{+0.10}_{-0.09}$&$10.70^{+1.19}_{-0.88}$&$5.69$\\ \noalign{\smallskip}
743&10 24 02.44&-57 44 36.05&Cl Westerlund    2     5          &O5/5.5V/III(f)     &$13.80$&$4.54^{+0.09}_{-0.07}$&$5.95^{+0.06}_{-0.08}$&$4.24^{+0.09}_{-0.08}$&$11.18^{+1.08}_{-0.81}$&$10.56$\\ \noalign{\smallskip}
770&10 24 06.64&-57 47 15.88&Cl* Westerlund    2    NRM       3&O9.5V              &$17.61$&$4.58^{+0.08}_{-0.08}$&$7.75^{+0.05}_{-0.07}$&$4.14^{+0.06}_{-0.05}$&$13.44^{+1.03}_{-0.98}$&$2.07$\\ \noalign{\smallskip}
789&10 24 16.25&-57 43 43.75&Cl* Westerlund    2    NRM       2&O8.5III            &$15.94$&$4.62^{+0.05}_{-0.07}$&$7.38^{+0.04}_{-0.05}$&$4.02^{+0.05}_{-0.05}$&$12.73^{+0.68}_{-0.87}$&$2.47$\\ \noalign{\smallskip}
793&10 24 18.40&-57 48 29.77&WR  20b                           &WN6ha              &$14.61$&$4.38^{+0.06}_{-0.05}$&$7.97^{+0.09}_{-0.10}$&$4.60^{+0.08}_{-0.07}$&$7.94^{+0.55}_{-0.43}$&$22.34$\\ \noalign{\smallskip}
797&10 24 21.29&-57 47 27.53&Cl* Westerlund    2    NRM       1&O6V                &$15.70$&$4.64^{+0.04}_{-0.06}$&$7.04^{+0.04}_{-0.04}$&$4.14^{+0.06}_{-0.05}$&$13.12^{+0.55}_{-0.76}$&$3.59$\\ \noalign{\smallskip}
822&10 24 39.20&-57 45 21.20&2MASS J10243919-5745211           &O5V                &$16.03$&$4.61^{+0.06}_{-0.08}$&$7.03^{+0.04}_{-0.05}$&$4.00^{+0.06}_{-0.05}$&$13.10^{+0.77}_{-0.93}$&$1.68$\\ \noalign{\smallskip}
895&10 25 47.07&-58 21 27.66&THA 35-II-41                      &HAeBe              &$13.55$&$4.56^{+0.09}_{-0.08}$&$4.14^{+0.05}_{-0.07}$&$4.78^{+0.15}_{-0.13}$&$13.31^{+1.08}_{-0.87}$&$4.89$\\ \noalign{\smallskip}
907&10 25 56.51&-57 48 43.54&WR  21a                           &WN+                &$13.62$&$4.37^{+0.05}_{-0.04}$&$6.34^{+0.09}_{-0.09}$&$4.45^{+0.09}_{-0.09}$&$8.59^{+0.49}_{-0.39}$&$42.02$\\ \noalign{\smallskip}

\hline
\end{tabular}}
\end{table*}

\subsubsection{Summary of results \label{summaryofresults}}

\begin{figure}
\begin{center}
\includegraphics[width=\columnwidth]{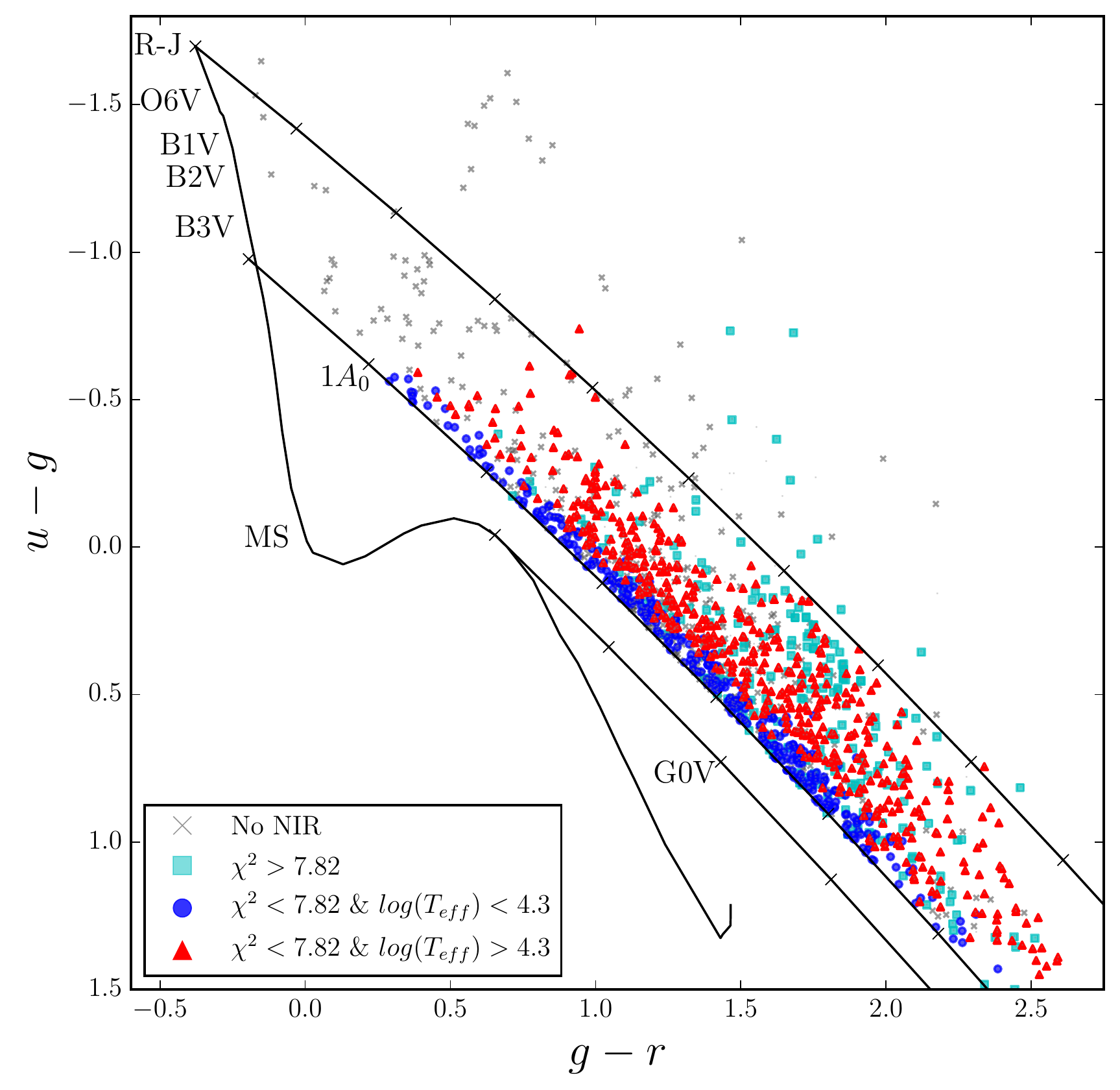}
\caption{$\ugr$ diagram showing the stages of selection. Red triangles are the final selection used for further discussion. All of the objects clearly above the RJ reddening vector are returned as bad fits.}
\label{fig:cc_chi}
\end{center}
\end{figure}

Figure \ref{fig:cc_chi} shows the stages in the selection process: first, those stars without a match to good quality NIR photometry have to be set aside (shown as grey crosses in the Figure); next, those with `poor' $\chi^2$ values (the cyan-coloured squares); finally the good fits are divided in two groups based on their effective temperature. Those with a median posterior effective temperature exceeding 20000K, or equivalently $\ltf \geq 4.3$, are shown as red triangles  while those that are assigned cooler fits are shown as blue squares. The hotter stars are our target group of spectral type B2 and earlier.

Counter-intuitively perhaps, it can be seen in Figure \ref{fig:cc_chi}
that in the domain where $g-r < 0.5$, only 12 stars could be matched
with good NIR photometry. This is because lowly reddened UV-excess objects
detected in VPHAS+ are commonly too faint for detection in 2MASS due to
their blue SEDs -- for instance, some of these objects
will be under-luminous hot compact objects. Unsurprisingly the cyan coloured squares representing objects with poor fits are frequently to be
found above the Rayleigh-Jeans limit -- only 2 objects with
accepted fits fall into this part of the diagram. It is
reassuring that there is some offset between the $R_V=3.1$ B3V
reddening vector, serving as lower bound to the selection region, and
the spread of hotter objects: it suggests that few, if any, stars
hotter than $\ltf = 4.3$ have been missed (given our other
constraints, such as the magnitude limits). It is worth noting that the selection of objects that occupied the 0.1\,mag wide band directly below the B3V reddening vector in $u-g$ provided just 1 star out of 374 with $\ltf \geq 4.3$ and $\chi^2<7.82$.

The main groupings emerging from the fitting process of all 1073 objects are shown in table \ref{table:breakdown}. All of the objects along with their photometry and derived parameters are tabulated in Tables \ref{table:main_photom} and \ref{table:main_params}.

\begin{table}

\caption{Breakdown of the number of new OB candidates, previously identified OB candidates and objects with known spectral type according to effective temperature and fit quality.} \label{table:breakdown}

\begin{center}

\begin{tabular}{ccc}

\hline \noalign{\smallskip}

\multicolumn{3}{c}{All Objects: 1073}\\ \noalign{\smallskip}

\hline \noalign{\smallskip}

$\ltf \geq 4.3$ & $\chi^2 \leq 7.82$ & $\chi^2 > 7.82$ \\ \noalign{\smallskip}

\hline \noalign{\smallskip}

Total & 527 & 145\\

New Candidate OBs & 489 & 98\\

Old Candidate OBs & 19 & 28\\

Known O - B2 stars & 19 & 10\\

Other & 0 & 1 C star \& 6 WR stars\\ \noalign{\smallskip}

\hline \noalign{\smallskip}

$\ltf < 4.3$ & $\chi^2 \leq 7.82$ & $\chi^2 > 7.82$ \\ \noalign{\smallskip}

\hline \noalign{\smallskip}

Total & 321 & 80\\

New Candidate OBs & 320 & 78\\

Old Candidate OBs & 0 & 2\\

Known O - B2 stars & 1 & 0\\ \noalign{\smallskip}

\hline \noalign{\smallskip}

All $\ltf$ & 848 & 225\\ \noalign{\smallskip}

\hline

\end{tabular}
\end{center}
\end{table}

\begin{sidewaystable*}

\caption{Sample table containing the positions and photometry of all 1073 objects. The first five columns are the objects IDs given in this study, VPHAS ID, \protect\cite{Moffatetal1991}, MSP ID, \protect\cite{Alvarezetal2013}, VA ID, \protect\cite{Tsujimotoetal2007}, TFT ID, and in SIMBAD, SIMBAD ID, where applicable. The full table can be found in the electronic version of this paper.} \label{table:main_photom}

\resizebox{\textwidth}{!}{
\begin{tabular}{ccccccccccccccccccccccccc}

\hline
ID & MSP ID & VA ID & TFT ID & SIMBAD ID & ST & RA (J2000) & DEC (J2000) & u & err u & g & err g & r & err r & i & err i & Ha & err Ha & J & err J & H & err H & K & err K\\
\hline
1&--&--&--&--&--&10 13 09.25&-58 01 58.10&$15.061$&$0.002$&$15.008$&$0.001$&$13.643$&$0.001$&$12.861$&$13.044$&$0.001$&$0.001$&$11.696$&$0.021$&$11.255$&$0.022$&$10.837$&$0.021$\\
52&--&--&--&--&--&10 14 40.36&-57 24 26.24&$16.937$&$0.006$&$15.487$&$0.001$&$12.959$&$0.001$&$11.468$&$12.345$&$0.001$&$0.001$&$8.693$&$0.024$&$7.870$&$0.036$&$7.414$&$0.027$\\
89&--&--&--&--&--&10 15 24.77&-57 44 09.28&$20.081$&$0.101$&$19.776$&$0.025$&$18.265$&$0.016$&$17.376$&$17.914$&$0.022$&$0.019$&$15.900$&$0.089$&$15.331$&$0.094$&$14.970$&$0.121$\\
162&--&--&--&--&--&10 16 31.33&-57 48 18.68&$19.199$&$0.034$&$18.923$&$0.010$&$17.323$&$0.006$&$16.565$&$17.392$&$0.017$&$0.007$&$15.246$&$0.054$&$14.763$&$0.079$&$14.509$&$0.096$\\
164&--&--&--&--&--&10 16 31.97&-57 56 02.37&$19.675$&$0.050$&$19.702$&$0.026$&$17.937$&$0.014$&$17.316$&$16.223$&$0.008$&$0.019$&$15.476$&$0.064$&$14.878$&$0.062$&$14.549$&$0.090$\\
175&--&--&--&--&--&10 16 42.53&-57 32 47.65&$17.559$&$0.009$&$16.168$&$0.002$&$13.576$&$0.001$&$12.078$&$13.053$&$0.001$&$0.001$&$9.327$&$0.026$&$8.474$&$0.049$&$7.953$&$0.027$\\
413&--&--&--&--&--&10 19 47.82&-57 50 38.64&$17.677$&$0.014$&$17.057$&$0.004$&$15.127$&$0.002$&$13.894$&$14.697$&$0.003$&$0.001$&$11.830$&$0.026$&$11.179$&$0.027$&$10.872$&$0.027$\\
496&--&--&--&--&--&10 21 20.56&-57 43 09.40&$15.521$&$0.003$&$15.231$&$0.001$&$13.666$&$0.001$&$12.692$&$13.284$&$0.001$&$0.001$&$11.014$&$0.023$&$10.524$&$0.023$&$10.292$&$0.021$\\
578&--&--&--&--&--&10 22 19.90&-57 46 11.21&$17.855$&$0.011$&$16.506$&$0.002$&$14.037$&$0.001$&$12.592$&$13.513$&$0.001$&$0.001$&$9.870$&$0.024$&$9.100$&$0.024$&$8.600$&$0.021$\\
601&--&--&--&--&--&10 22 35.02&-58 33 37.82&$16.554$&$0.005$&$16.692$&$0.002$&$15.716$&$0.002$&$15.153$&$15.471$&$0.004$&$0.002$&$14.220$&$0.056$&$13.883$&$0.066$&$13.810$&$0.064$\\
677&182    &178&112&Cl* Westerlund    2    MSP     182&O4V-III((f))      &10 23 56.18&-57 45 30.00&$15.587$&$0.004$&$15.349$&$0.002$&$13.624$&$0.001$&$12.613$&$13.211$&$0.001$&$0.001$&$10.520$&$0.015$&$10.050$&$0.008$&$9.750$&$0.020$\\
712&157    &584&--&2MASS J10240073-5745253           &O8V               &10 24 00.76&-57 45 25.65&$15.376$&$0.004$&$15.137$&$0.002$&$13.450$&$0.001$&$12.437$&$13.044$&$0.001$&$0.001$&$11.660$&$0.038$&$11.150$&$0.027$&$10.790$&$0.031$\\
724&263    &722&202&Cl* Westerlund    2    MSP     263&O6V               &10 24 01.52&-57 45 57.00&$16.696$&$0.008$&$16.094$&$0.003$&$14.055$&$0.001$&$12.859$&$13.633$&$0.002$&$0.001$&$10.520$&$0.025$&$9.910$&$0.035$&$9.530$&$0.036$\\
732&167    &804&217&Cl* Westerlund    2    MSP     167&O8V               &10 24 02.04&-57 45 27.94&$15.858$&$0.005$&$15.398$&$0.002$&$13.430$&$0.001$&$12.306$&$13.007$&$0.001$&$0.001$&$9.960$&$0.122$&$9.357$&$0.158$&$8.982$&$0.095$\\
737&203/444&857&224&Cl* Westerlund    2    NRM       4&O4.5V             &10 24 02.29&-57 45 35.26&$14.604$&$0.003$&$14.310$&$0.001$&$12.566$&$0.001$&$11.560$&$12.170$&$0.001$&$0.001$&$9.450$&$0.063$&$8.950$&$0.090$&$8.520$&$0.050$\\
763&171    &1039&298&Cl* Westerlund    2    MSP     171&O4-5V             &10 24 04.90&-57 45 28.35&$15.886$&$0.005$&$15.470$&$0.002$&$13.616$&$0.001$&$12.552$&$13.206$&$0.001$&$0.001$&$10.480$&$0.018$&$10.000$&$0.039$&$9.740$&$0.033$\\
770&383    &--&314&Cl* Westerlund    2    NRM       3&O9.5V             &10 24 06.64&-57 47 15.88&$18.263$&$0.023$&$17.608$&$0.006$&$15.502$&$0.003$&$14.248$&$14.952$&$0.003$&$0.002$&$11.736$&$0.028$&$11.008$&$0.025$&$10.536$&$0.023$\\
789&--&--&388&Cl* Westerlund    2    NRM       2&O8.5III           &10 24 16.25&-57 43 43.75&$16.495$&$0.007$&$15.936$&$0.003$&$13.883$&$0.001$&$12.664$&$13.432$&$0.001$&$0.001$&$10.316$&$0.023$&$9.624$&$0.023$&$9.236$&$0.021$\\
797&--&--&405&Cl* Westerlund    2    NRM       1&O6V               &10 24 21.29&-57 47 27.53&$16.048$&$0.006$&$15.703$&$0.002$&$13.796$&$0.001$&$12.636$&$13.367$&$0.001$&$0.001$&$10.435$&$0.024$&$9.745$&$0.022$&$9.386$&$0.019$\\
822&--&--&447&2MASS J10243919-5745211           &O5V               &10 24 39.20&-57 45 21.20&$16.519$&$0.007$&$16.033$&$0.003$&$14.084$&$0.001$&$12.919$&$13.644$&$0.002$&$0.001$&$10.709$&$0.026$&$10.060$&$0.027$&$9.679$&$0.023$\\
\hline
\end{tabular}

}

\caption{Sample table containing the derived parameters of all 1073 objects. The $16^{\rm th}$, $50^{\rm th}$ and $84^{\rm th}$ percentiles are given for each parameter as well as the $\chi^2$ value at the $50^{\rm th}$ percentile. The notes column indicates if the object shows emission (EM), is a sub-luminous candidate (SUB), is a blue supergiant candidate (BSG) or is a new O star candidate near Wd 2 (WD2) with similar reddening. The full table can be found in the electronic version of this paper.}\label{table:main_params}

\resizebox{\textwidth}{!}{
\begin{tabular}{cccccccccccccccc}

\hline
ID & $\ltf$ $\rm P16^{th}$ & $\ltf$  $\rm P50^{th}$ & $\ltf$ $\rm P84^{th}$ & A0 $\rm P16^{th}$ &  A0 $\rm P50^{th}$ & A0 $\rm P84^{th}$ & RV $\rm P16^{th}$ & RV $\rm P50^{th}$ & RV $\rm P84^{th}$ & DM $\rm P16^{th}$ & DM $\rm P50^{th}$ & DM $\rm P84^{th}$ & $\chi^2$ & Notes\\
\hline
1&$4.39$&$4.44$&$4.39$&$4.54$&$4.62$&$4.54$&$3.65$&$3.73$&$3.65$&$12.14$&$12.65$&$12.14$&$37.82$&EM    \\
52&$4.31$&$4.36$&$4.31$&$8.29$&$8.38$&$8.29$&$3.72$&$3.77$&$3.72$&$7.54$&$7.93$&$7.54$&$2.28$&BSG   \\
89&$4.33$&$4.43$&$4.33$&$5.09$&$5.26$&$5.09$&$3.67$&$3.79$&$3.67$&$15.69$&$16.57$&$15.69$&$1.97$&SUB   \\
162&$4.40$&$4.47$&$4.40$&$4.69$&$4.80$&$4.69$&$3.27$&$3.35$&$3.27$&$15.89$&$16.58$&$15.89$&$5.05$&SUB   \\
164&$4.46$&$4.54$&$4.46$&$5.64$&$5.73$&$5.64$&$4.03$&$4.13$&$4.03$&$16.39$&$17.23$&$16.39$&$39.93$&EM    \\
175&$4.37$&$4.43$&$4.37$&$8.48$&$8.59$&$8.48$&$3.72$&$3.77$&$3.72$&$8.59$&$9.15$&$8.59$&$1.04$&BSG   \\
413&$4.52$&$4.61$&$4.52$&$6.77$&$6.83$&$6.77$&$3.72$&$3.77$&$3.72$&$13.19$&$14.23$&$13.19$&$6.66$&WD2   \\
496&$4.46$&$4.54$&$4.46$&$5.53$&$5.62$&$5.53$&$3.69$&$3.75$&$3.69$&$12.08$&$12.99$&$12.08$&$4.61$&--&\\
578&$4.32$&$4.37$&$4.32$&$8.12$&$8.22$&$8.12$&$3.75$&$3.80$&$3.75$&$8.83$&$9.23$&$8.83$&$2.32$&BSG   \\
601&$4.32$&$4.36$&$4.32$&$3.35$&$3.45$&$3.35$&$3.57$&$3.69$&$3.57$&$14.46$&$14.83$&$14.46$&$0.97$&SUB   \\
677&$4.56$&$4.63$&$4.56$&$6.31$&$6.35$&$6.31$&$3.97$&$4.03$&$3.97$&$12.56$&$13.38$&$12.56$&$24.78$&--&\\
712&$4.61$&$4.66$&$4.61$&$4.56$&$4.60$&$4.56$&$2.86$&$2.91$&$2.86$&$14.60$&$15.24$&$14.60$&$72.87$&--&\\
724&$4.53$&$4.61$&$4.53$&$7.19$&$7.25$&$7.19$&$3.89$&$3.94$&$3.89$&$11.84$&$12.79$&$11.84$&$7.64$&--&\\
732&$4.52$&$4.60$&$4.52$&$7.03$&$7.14$&$7.03$&$4.03$&$4.11$&$4.03$&$11.18$&$12.14$&$11.18$&$1.89$&--&\\
737&$4.48$&$4.56$&$4.48$&$6.43$&$6.51$&$6.43$&$4.10$&$4.17$&$4.10$&$10.39$&$11.30$&$10.39$&$2.52$&--&\\
763&$4.55$&$4.62$&$4.55$&$6.39$&$6.44$&$6.39$&$3.74$&$3.80$&$3.74$&$12.40$&$13.26$&$12.40$&$18.43$&--&\\
770&$4.50$&$4.58$&$4.50$&$7.68$&$7.75$&$7.68$&$4.09$&$4.14$&$4.09$&$12.46$&$13.44$&$12.46$&$2.07$&--&\\
789&$4.55$&$4.62$&$4.55$&$7.33$&$7.38$&$7.33$&$3.98$&$4.02$&$3.98$&$11.85$&$12.73$&$11.85$&$2.47$&--&\\
797&$4.58$&$4.64$&$4.58$&$7.00$&$7.04$&$7.00$&$4.09$&$4.14$&$4.09$&$12.36$&$13.12$&$12.36$&$3.59$&--&\\
822&$4.54$&$4.61$&$4.54$&$6.98$&$7.03$&$6.98$&$3.95$&$4.00$&$3.95$&$12.16$&$13.10$&$12.16$&$1.68$&--&\\

\hline
\end{tabular}
}

\end{sidewaystable*}

\subsubsection{Contaminants}

\begin{figure}
\begin{center}
\includegraphics[width=\columnwidth]{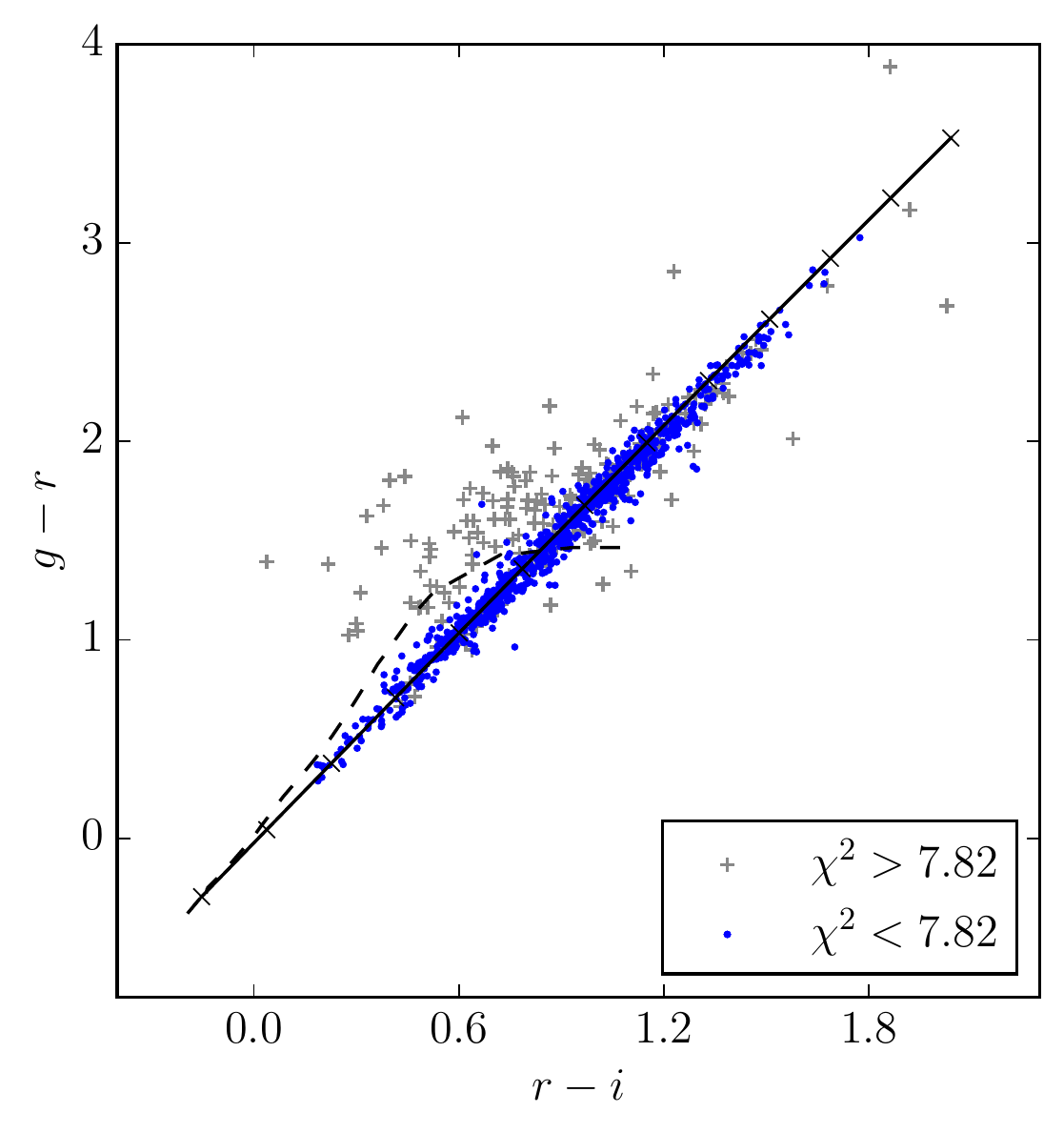}
\caption{Positions of objects with $\chi^2 \leq 7.82$ (blue dots) and $\chi^2 > 7.82$ (grey crosses) in the $(r-i, g-r)$ plane. The solid black line is the reddening vector of an O9V with $R_V=3.8$. The dashed line is the unreddened main sequence. We find that a large number of objects with `poor' fits fall away from the OB star reddening vector. These objects show colours that are consistent with eclipsing W UMa contact binaries.}
\label{fig:binaries}
\end{center}
\end{figure}

The $(\chi^2 > 7.82)$ fits have a range of causes. The most frequent are likely to be contact binaries or the products of poor photometry.

Contact binaries may find their way into the selection because they are both quite common and rapidly variable. Figure \ref{fig:binaries} shows how around half of the $\chi^2 \leq 7.82$ objects clearly separate in the $(r-i, g-r)$ colour-colour diagram away from the OB stars towards redder $g-r$ at fixed $r-i$. This is plausibly the signature of contact binary (W UMa) interlopers. W UMa systems are doubly eclipsing binaries in which the brightness in any one band scarcely remains constant over time. These objects have typical orbital periods of 8 hours with two pronounced minima per cycle \citep{rucinski1992}. The u/g/r VST exposures are taken sequentially with about 15 minutes elapsing between u and g, and g and r. If the g band exposure of a W UMa system is taken at or near minimum light, its measured $u-g$ colour is bluer than true, while $g - r$ is redder, potentially pushing the star up into our OB selection. However these objects fail to pass as OB stars when the whole OnIR SED fit is performed, hence their poor $\chi^2$ values.
It has been estimated that there is around 1 W UMa system for every $\sim130$ main sequence stars \citep{rucinski1992}. So finding perhaps as many as $\sim100$ in our OB selection, given $\sim100000$ stars across the 2 square degrees with $u/g/r$ photometry, is reasonable.

The second common origin for the poor fits is likely due to photometry affected by blending or incorrect cross-matching between bands. In the crowded core of Wd 2 this is an obvious difficulty (see Figures 4 and 10).

The literature search already reported in section \ref{summaryofresults} revealed that high $\chi^2$ may be linked to extreme objects like WR stars (6 examples) and carbon stars (1 only). Another rare contaminant may be white dwarf/M dwarf binaries that can present blue $u-g$, alongside red $r-i$. The blue white-dwarf light begins to be overwhelmed by the red dwarf's light with increasing wavelength, shifting the combined colours below and to the right of the OB reddening track in the $(g-r, r-i)$ diagram (fig~\ref{fig:binaries}). Such objects are known to co-locate with reddened OB stars in the $(u-g, g-r)$ diagram or they may fall beyond the RJ reddening vector \citep{Smolcicetal2004}.

\subsection{Parameters of the candidate OB stars}

\begin{figure}
\begin{center}
\includegraphics[width=\columnwidth]{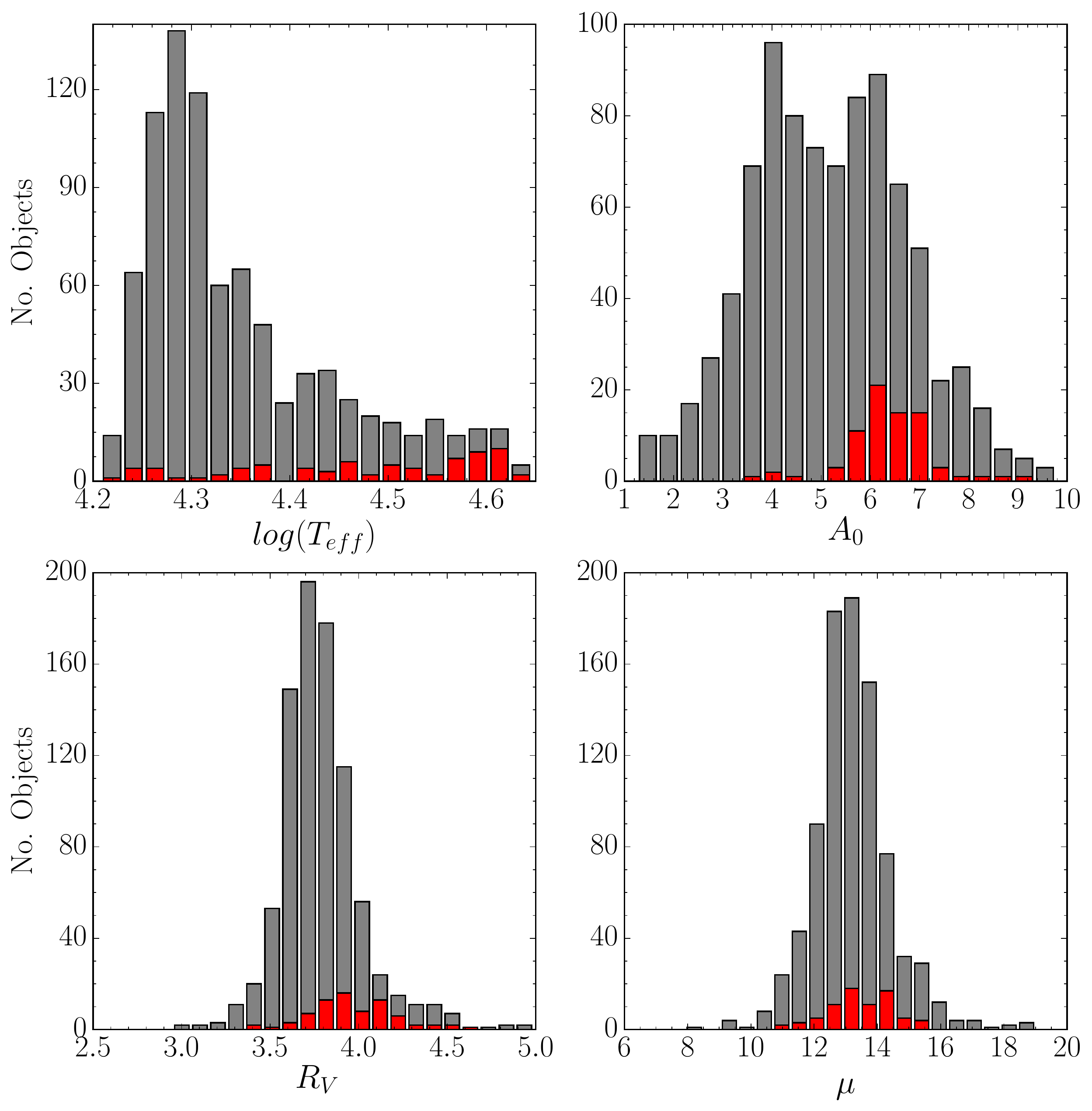}
\caption{Distribution of the best fit parameters for the selection of objects with $\chi^2<7.82$. The red bars are objects within and 8 arcmin box of Wd 2 while the grey bars are the wider selection. We find that the objects spatially associated with Wd 2 show a tight distribution in $A_0$ and provide an over density of objects in the $5.5 \leq A_0 \leq 7$ range and also show a wider spread in $R_V$.}
\label{fig:param_hist}
\end{center}
\end{figure}

Figure \ref{fig:param_hist} shows the distribution of stellar
parameters across the entire selection for the objects fitting successfully to a reddened OB-star SED ($\chi^2 \leq 7.82$ and $\ltf \geq 4.3$). Coloured in red are the results for all
objects within an 8 arcmin box centred on Wd 2 (drawn in
Figure \ref{fig:area}). It can be seen that those objects in or near
the cluster are reported to have similar extinction in the range $5.5 \leq A_0 \leq 7$ (top right
panel in Figure \ref{fig:param_hist}).
Otherwise, the reddenings range more broadly across the full 2 square
degrees from $A_0 \simeq 3$ up to $A_0 \simeq 8$.  Other
features of this particular sight-line are that larger than standard $R_V$
is favoured -- a roughly normal distribution in $R_V$ about a mean
value of $R_V=3.84\pm0.25$ is obtained -- and that most of the
selected stars are attributed distances of between $\sim 2$~kpc ($\mu \simeq 11$) and $\sim 6$~kpc ($\mu \simeq 14$). The objects in/near Wd2
tend toward the higher end of the distance modulus range and show a fairly wide spread in extinction law with $3.5 \leq R_V \leq 4.5$.

Echoing the initial mass function (IMF), the distribution
in median $\ltf$ values is heavily skewed towards the lower end. The turn over in the $\ltf$ distribution at just below $\ltf= 4.3$ further supports the conclusion that our initial selection
of VPHAS+ sources in the $\ugr$ diagram is essentially complete in the
desired O to B2 effective temperature range (given our magnitude limits). The coolest object in the candidate list is $\sim$16000~K.

Stars with median estimated effective temperatures in excess of 30000K ($\ltf \geq 4.477$) are regarded as candidate O stars. Of the new discoveries 74 meet this criterion. We can further subdivide this group to distinguish the highly probable O stars: 28 objects have a $16^{\rm th}$ percentile $\ltf$ exceeding 4.477. Seven of these may be sdO stars (see section \ref{sec:lum}).

Predictably, many of the hottest candidates are in and around Wd 2: this young massive
cluster does indeed stand out in this part of the Galactic Plane. Moreover the top left panel in \ref{fig:param_hist} suggests a relative lack of cooler OB stars within the 8 arcmin box centred on the cluster. This could be taken to imply that the stellar mass function of Wd 2 and environs is top heavy. At the same time, there are biases that can favour the detection of more massive stars at the likely distance of the cluster ($\mu \sim 13 - 14 $) -- namely, the effects of crowding (less massive fainter stars are more likely to be lost in blends) and of magnitude limited selection. However this is unlikely to be all of the explanation given that there are plenty of examples of $A_0 \sim 5.5 - 7$ cool candidates with a similar estimated distance modulus.

\begin{figure}
\begin{center}
\includegraphics[width=\columnwidth]{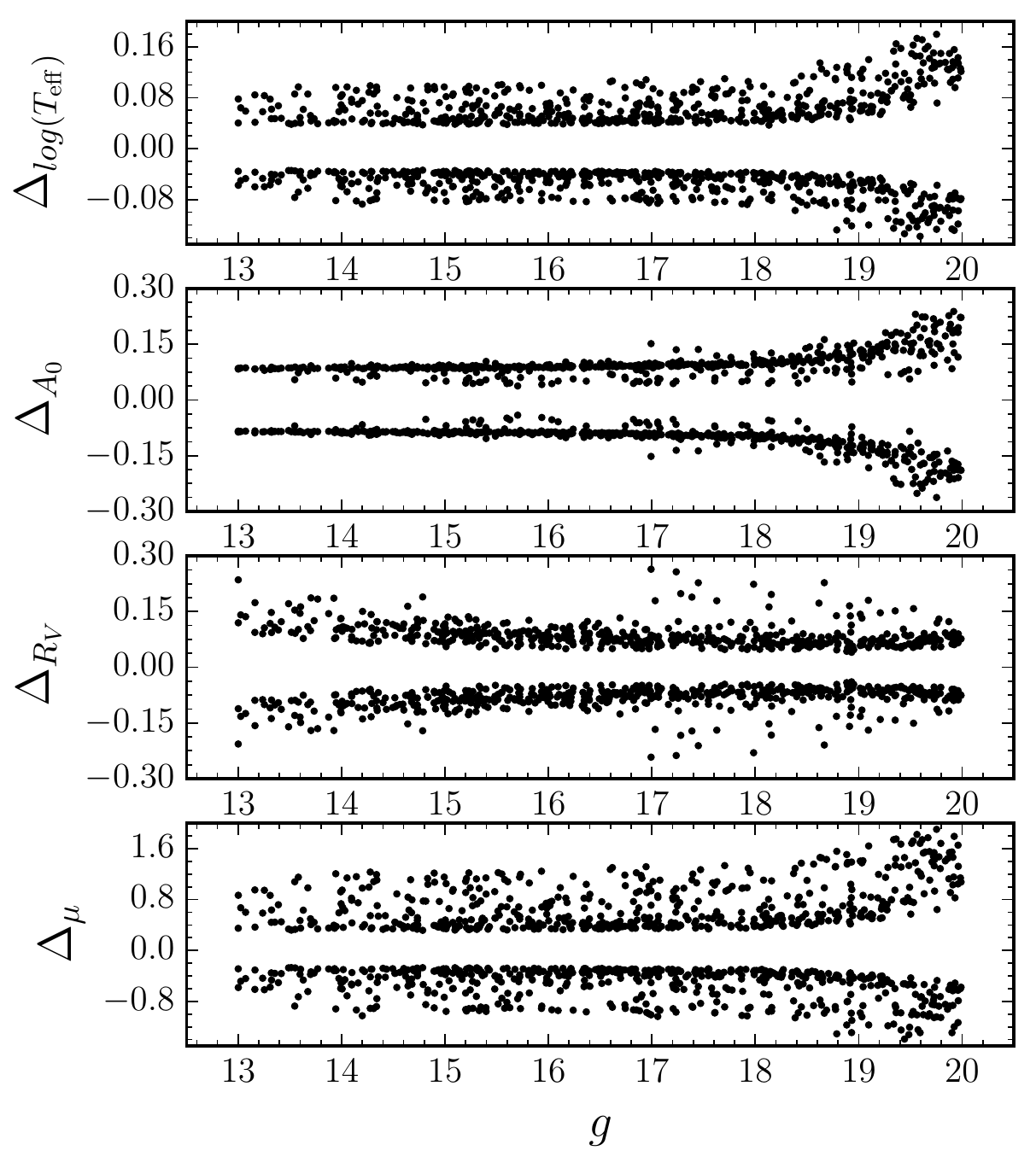}
\caption{Uncertainty on each parameter as a function of g band magnitude. Uncertainties are derived from the $16^{\rm th}$ and $84^{\rm th}$ percentiles of the posterior distributions.}
\label{fig:param_errs}
\end{center}
\end{figure}

Figure \ref{fig:param_errs} shows the upper and lower
uncertainties on each parameter as a function of $g$-band magnitude
for all $\chi^2<7.82$ objects. The uncertainty on $\ltf$ and
$A_0$ increases for fainter objects, tracking the increase with rising magnitude of the photometric errors. Conversely the uncertainty on
$R_V$ shows a slight increase with decreasing magnitude at the bright
end. $R_V$ is more difficult to determine for bright objects as they tend to be less obscured. Nevertheless it is evident that
both $R_V$ and $A_0$ are consistently well determined across the
entire magnitude range. Our OnIR SED fits deliver $A_0$ to within
$\lesssim0.09$\,mag up to $18^{th}$ magnitude, rising up to
$\lesssim0.25$\,mag at $20^{th}$ magnitude. We find the median
uncertainty on $R_V$ to be $0.081$.

\subsection{Inferences from the best-fit parameters and other aspects
  of the photometry \label{sec:lum}}

\begin{figure*}
\begin{center}
\includegraphics[width=\textwidth]{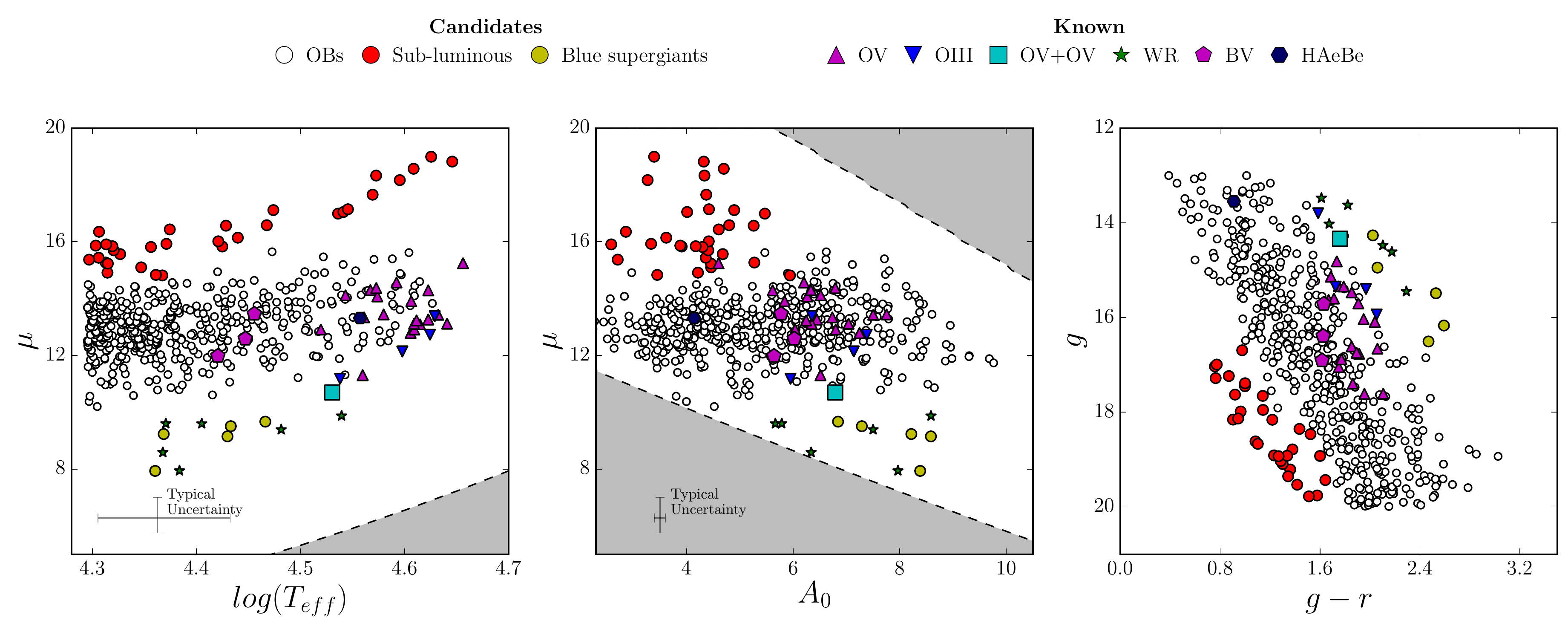}
\caption{2-D distribution of the best fit parameters for the final
  selection of OB candidates ($\chi^2 \leq 7.82$ and $\ltf \geq 4.3$) and objects in the selection with known spectral type (the carbon star lays outside the range of 2 of these diagrams and was therefore not included). Objects shown in red and yellow are thought to be candidate sub-luminous OB stars and candidate blue supergiants respectively. The areas shaded in grey are where we
cannot detect OB stars given the survey limits.}
\label{fig:param_plot}
\end{center}
\end{figure*}

A richer understanding of the candidate objects can be obtained from a
combination of more scrutiny of the fit parameters obtained and from a
fuller utilisation of the VPHAS+ photometry at our disposal.  So far
the focus has been on the information to be extracted from the
individual OnIR SEDs -- treating all candidates as if they are well
described as reddened, single, main-sequence OB stars.  We can learn
more through consideration of the ensemble of objects, and if use is
made of the narrowband H$\alpha$ band to separate out emission line
stars.

First we acknowledge and relax the main sequence assumption applied so
far.  The first two panels of Figure \ref{fig:param_plot} show scatter
plots of the best-fit median distance modulus, $\mu$,
vs. $\ltf$ and vs. $A_0$ for the candidate OB stars.
Different symbols are over-plotted to pick out the already known
objects listed in SIMBAD as well as the O stars of \cite{Alvarezetal2013}. The areas shaded in grey are where we cannot detect OB stars given
the survey limits.   The objects plotted as red circles have
relatively low extinction but, if we take the returned distance moduli at
face value, they would have to be construed as very distant ($>
10$~kpc) when compared to the known OB stars.  It is more plausible
that these are intrinsically sub-luminous objects rather than distant
OB stars located in remarkably clear reddening holes. Their scattered spatial distribution across the whole field shown in Figure \ref{fig:area} supports this argument.

The converse argument can be applied to those objects plotted as
yellow circles: they are found to have more than 6 magnitudes of
extinction but are seemingly very close (less than $\sim$700pc away, $\mu <9$).  We  suspect that these objects are intrinsically much
higher-luminosity, evolved B stars. The proximity of these stars in
the figures to the (poorly-fit) known WR stars, including the
highly-luminous WR20a, lends credibility to this interpretation.

Referring back to the photometry in the form of a ($g, g-r$) colour
magnitude diagram (CMD), these interpretations are seen to make sense
-- the sub-luminous and over-luminous objects form tracks separated from the main-sequence -- see the third panel of Figure \ref{fig:param_plot}.  Table
\ref{table:sub/over-luminous} lists these extreme objects.  There
will be further discussion of them in Section 6. The main concentration of objects appears in the $11.5 < \mu < 14$\,mag range which equates to distances ranging from $2 \-- 6$kpc. This encloses the derived distance range of the Carina arm traced in CO by \cite{Grabelsky1988}, near its tangent.

\begin{table*}
\caption{Table containing the derived stellar parameters of the sub-luminous and blue supergiant candidates in the $\chi^2 \leq 7.82$ and $\ltf \geq 4.3$ group.}
\label{table:sub/over-luminous}
\begin{tabular}{ccccccccccc}

\hline\noalign{\smallskip}
ID&RA&DEC&g&$log(T_{eff})$&$A_0$&$R_V$&$\mu$&$\chi^2$\\
\hline\noalign{\smallskip}
\multicolumn{9}{c}{Sub-Luminous} \\

\hline\noalign{\smallskip}
19&10 13 40.87&-57 43 15.38&$18.13$&$4.65^{+0.04}_{-0.05}$&$4.32^{+0.09}_{-0.09}$&$4.95^{+0.16}_{-0.15}$&$18.82^{+0.50}_{-0.69}$&$7.09$\\ [0.2cm]
28&10 13 54.74&-58 14 33.73&$18.91$&$4.31^{+0.05}_{-0.04}$&$4.36^{+0.13}_{-0.13}$&$3.82^{+0.17}_{-0.16}$&$15.44^{+0.45}_{-0.37}$&$3.53$\\ [0.2cm]
82&10 15 21.46&-57 53 30.64&$19.43$&$4.37^{+0.08}_{-0.06}$&$5.94^{+0.13}_{-0.13}$&$4.03^{+0.10}_{-0.10}$&$14.82^{+0.77}_{-0.52}$&$1.11$\\ [0.2cm]
86&10 15 22.76&-57 48 21.73&$18.67$&$4.57^{+0.08}_{-0.08}$&$4.34^{+0.16}_{-0.17}$&$4.17^{+0.23}_{-0.21}$&$18.33^{+0.98}_{-0.99}$&$2.10$\\ [0.2cm]
89&10 15 24.77&-57 44 09.28&$19.78$&$4.43^{+0.13}_{-0.09}$&$5.26^{+0.15}_{-0.17}$&$3.79^{+0.13}_{-0.12}$&$16.57^{+1.47}_{-0.88}$&$1.97$\\ [0.2cm]
96&10 15 36.78&-57 46 58.03&$17.65$&$4.37^{+0.05}_{-0.05}$&$3.34^{+0.10}_{-0.10}$&$3.13^{+0.11}_{-0.10}$&$15.93^{+0.53}_{-0.42}$&$0.82$\\ [0.2cm]
153&10 16 25.54&-58 33 18.54&$18.93$&$4.61^{+0.06}_{-0.08}$&$4.70^{+0.09}_{-0.09}$&$3.84^{+0.11}_{-0.11}$&$18.57^{+0.83}_{-0.99}$&$6.51$\\ [0.2cm]
162&10 16 31.33&-57 48 18.68&$18.92$&$4.47^{+0.09}_{-0.07}$&$4.80^{+0.10}_{-0.11}$&$3.35^{+0.09}_{-0.09}$&$16.58^{+1.04}_{-0.69}$&$5.05$\\ [0.2cm]
177&10 16 44.32&-58 01 29.97&$19.02$&$4.32^{+0.06}_{-0.05}$&$4.17^{+0.13}_{-0.13}$&$3.60^{+0.14}_{-0.13}$&$15.84^{+0.54}_{-0.42}$&$2.71$\\ [0.2cm]
202&10 17 01.28&-58 05 29.31&$19.35$&$4.47^{+0.11}_{-0.08}$&$4.89^{+0.13}_{-0.14}$&$3.87^{+0.15}_{-0.14}$&$17.11^{+1.34}_{-0.86}$&$3.96$\\ [0.2cm]
219&10 17 15.78&-57 23 13.80&$19.53$&$4.37^{+0.07}_{-0.06}$&$4.60^{+0.15}_{-0.15}$&$3.57^{+0.16}_{-0.15}$&$16.43^{+0.75}_{-0.54}$&$1.31$\\ [0.2cm]
274&10 17 59.10&-58 01 05.92&$18.16$&$4.31^{+0.05}_{-0.04}$&$2.86^{+0.13}_{-0.13}$&$3.27^{+0.20}_{-0.18}$&$16.35^{+0.43}_{-0.36}$&$3.30$\\ [0.2cm]
275&10 17 59.67&-57 48 12.93&$17.03$&$4.30^{+0.04}_{-0.04}$&$2.71^{+0.10}_{-0.10}$&$3.61^{+0.18}_{-0.17}$&$15.37^{+0.35}_{-0.30}$&$1.17$\\ [0.2cm]
292&10 18 11.80&-58 20 12.24&$19.21$&$4.33^{+0.05}_{-0.05}$&$4.68^{+0.13}_{-0.13}$&$3.82^{+0.15}_{-0.14}$&$15.57^{+0.50}_{-0.41}$&$1.62$\\ [0.2cm]
402&10 19 39.37&-58 29 18.43&$19.76$&$4.31^{+0.07}_{-0.06}$&$5.27^{+0.15}_{-0.15}$&$3.72^{+0.12}_{-0.11}$&$15.27^{+0.64}_{-0.47}$&$3.79$\\ [0.2cm]
468&10 20 48.03&-57 45 45.52&$18.35$&$4.35^{+0.05}_{-0.05}$&$4.46^{+0.12}_{-0.12}$&$3.47^{+0.12}_{-0.12}$&$15.10^{+0.51}_{-0.41}$&$0.35$\\ [0.2cm]
472&10 20 53.73&-57 58 41.78&$17.95$&$4.55^{+0.10}_{-0.08}$&$4.42^{+0.08}_{-0.09}$&$3.94^{+0.12}_{-0.11}$&$17.14^{+1.22}_{-0.89}$&$3.28$\\ [0.2cm]
474&10 20 54.07&-58 02 32.59&$18.92$&$4.30^{+0.05}_{-0.04}$&$3.88^{+0.12}_{-0.12}$&$3.40^{+0.14}_{-0.13}$&$15.87^{+0.42}_{-0.35}$&$6.59$\\ [0.2cm]
484&10 21 07.91&-57 33 52.74&$18.16$&$4.57^{+0.09}_{-0.08}$&$4.37^{+0.08}_{-0.09}$&$3.70^{+0.11}_{-0.10}$&$17.66^{+1.09}_{-0.95}$&$3.39$\\ [0.2cm]
486&10 21 10.28&-58 10 46.02&$18.62$&$4.36^{+0.05}_{-0.05}$&$4.30^{+0.11}_{-0.11}$&$4.33^{+0.17}_{-0.16}$&$15.82^{+0.50}_{-0.43}$&$1.42$\\ [0.2cm]
601&10 22 35.02&-58 33 37.82&$16.69$&$4.36^{+0.05}_{-0.04}$&$3.45^{+0.10}_{-0.10}$&$3.69^{+0.13}_{-0.13}$&$14.83^{+0.46}_{-0.38}$&$0.97$\\ [0.2cm]
647&10 23 27.84&-57 54 56.79&$19.03$&$4.54^{+0.11}_{-0.09}$&$5.47^{+0.09}_{-0.10}$&$4.39^{+0.13}_{-0.12}$&$16.99^{+1.38}_{-1.04}$&$5.40$\\ [0.2cm]
810&10 24 31.14&-57 33 45.22&$16.99$&$4.60^{+0.07}_{-0.08}$&$3.27^{+0.15}_{-0.15}$&$4.03^{+0.26}_{-0.24}$&$18.17^{+0.94}_{-1.01}$&$1.71$\\ [0.2cm]
819&10 24 36.96&-58 22 47.18&$17.28$&$4.31^{+0.04}_{-0.04}$&$2.59^{+0.11}_{-0.11}$&$3.51^{+0.20}_{-0.18}$&$15.91^{+0.39}_{-0.33}$&$5.83$\\ [0.2cm]
842&10 24 58.98&-57 59 56.24&$17.45$&$4.42^{+0.07}_{-0.05}$&$3.91^{+0.14}_{-0.14}$&$4.11^{+0.23}_{-0.21}$&$15.83^{+0.77}_{-0.57}$&$0.66$\\ [0.2cm]
870&10 25 25.08&-57 59 04.50&$17.98$&$4.31^{+0.04}_{-0.04}$&$4.21^{+0.12}_{-0.12}$&$4.99^{+0.22}_{-0.23}$&$14.91^{+0.42}_{-0.33}$&$0.71$\\ [0.2cm]
894&10 25 47.01&-57 46 51.19&$19.09$&$4.32^{+0.06}_{-0.05}$&$4.41^{+0.15}_{-0.15}$&$3.84^{+0.18}_{-0.17}$&$15.70^{+0.54}_{-0.43}$&$1.49$\\ [0.2cm]
956&10 26 57.82&-57 36 16.66&$17.39$&$4.54^{+0.09}_{-0.08}$&$4.01^{+0.10}_{-0.10}$&$4.44^{+0.19}_{-0.17}$&$17.05^{+1.11}_{-0.88}$&$7.77$\\ [0.2cm]
989&10 27 34.13&-57 36 25.34&$18.79$&$4.31^{+0.05}_{-0.04}$&$4.46^{+0.13}_{-0.13}$&$3.59^{+0.14}_{-0.13}$&$15.23^{+0.46}_{-0.39}$&$3.04$\\ [0.2cm]
1024&10 28 14.60&-57 41 36.33&$17.24$&$4.44^{+0.07}_{-0.06}$&$3.62^{+0.14}_{-0.14}$&$4.22^{+0.26}_{-0.24}$&$16.14^{+0.82}_{-0.61}$&$1.70$\\ [0.2cm]
1052&10 28 50.85&-57 45 56.73&$17.63$&$4.63^{+0.05}_{-0.06}$&$3.39^{+0.12}_{-0.12}$&$3.80^{+0.18}_{-0.17}$&$18.99^{+0.61}_{-0.81}$&$2.77$\\ [0.2cm]
1059&10 28 56.00&-58 09 02.55&$18.46$&$4.42^{+0.07}_{-0.05}$&$4.42^{+0.10}_{-0.10}$&$3.27^{+0.09}_{-0.09}$&$16.02^{+0.72}_{-0.53}$&$5.96$\\ [0.2cm]
\hline\noalign{\smallskip}
\multicolumn{9}{c}{Blue supergiants} \\
\hline\noalign{\smallskip}
52&10 14 40.36&-57 24 26.24&$15.49$&$4.36^{+0.05}_{-0.05}$&$8.38^{+0.10}_{-0.10}$&$3.77^{+0.05}_{-0.05}$&$7.93^{+0.51}_{-0.40}$&$2.28$\\ [0.2cm]
175&10 16 42.53&-57 32 47.65&$16.17$&$4.43^{+0.08}_{-0.06}$&$8.59^{+0.10}_{-0.10}$&$3.77^{+0.05}_{-0.05}$&$9.15^{+0.82}_{-0.56}$&$1.04$\\ [0.2cm]
188&10 16 53.91&-57 55 02.11&$14.26$&$4.47^{+0.09}_{-0.06}$&$6.84^{+0.09}_{-0.09}$&$3.81^{+0.06}_{-0.06}$&$9.67^{+0.99}_{-0.63}$&$0.25$\\ [0.2cm]
452&10 20 31.60&-58 03 08.72&$14.95$&$4.43^{+0.07}_{-0.05}$&$7.29^{+0.09}_{-0.10}$&$4.03^{+0.07}_{-0.07}$&$9.51^{+0.77}_{-0.53}$&$0.91$\\ [0.2cm]
578&10 22 19.90&-57 46 11.21&$16.51$&$4.37^{+0.06}_{-0.05}$&$8.22^{+0.10}_{-0.10}$&$3.80^{+0.05}_{-0.05}$&$9.23^{+0.54}_{-0.41}$&$2.32$\\ [0.2cm]
\hline
\end{tabular}
\end{table*}

\begin{figure}
\begin{center}
\includegraphics[width=\columnwidth]{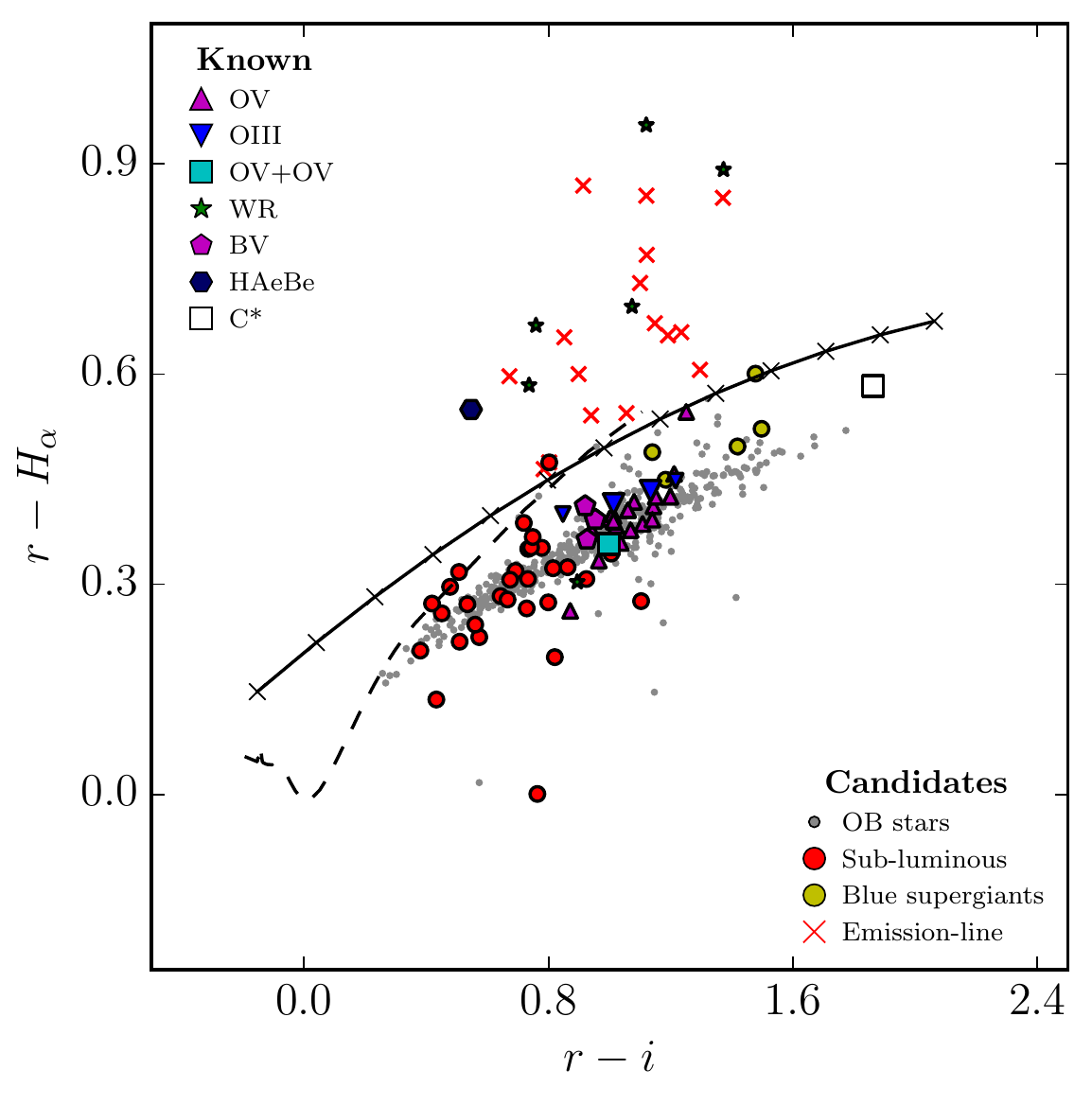}
\caption{13 of the $\chi^2 \leq 7.82$ and $\ltf \geq 4.3$ objects show H$\alpha$ excess. As emission is usually associated with circumstellar dust; the derived extinction may be incorrect. The solid line is the reddening vector of an O9V raised by 0.1 in $r - \rm H\alpha$.}
\label{fig:emission}
\end{center}
\end{figure}

We can also use the VPHAS+ H$\alpha$ measurements to uncover any
emission line stars in our selection. The presence of emission lines
implies the presence of ionized circumstellar gas which, among massive
OB stars, most commonly indicates classical Be stars with
circumstellar disks. Although the OnIR SEDs of classical Be stars are
not greatly different from normal B stars of similar effective
temperature, the derived interstellar extinctions from SED fits that
do not take into account the circumstellar continuum emission
will nevertheless be overestimated. We have used the $(r-i,r- \rm H\alpha)$
diagram to select all objects that lie more than $0.1$\,mag in
$r- \rm H\alpha$ above the O9V reddening vector (this equates to $\sim
10\AA$ in emission line equivalent width). Figure \ref{fig:emission}
shows this selection. Using the relation between $EW(\rm H\alpha)$ and
added colour excess $E(B-V)$ due to the presence of a circumstellar
disk in classical Be stars from \cite{Dachsetal1988}, we can estimate
that the derived reddenings ($A_0$) for our H$\alpha$-excess stars
will have been inflated by between $\sim$0.1 and $\sim$0.3
magnitudes. There are 17 of these objects in the $\chi^2 \leq 7.82$ and $\ltf \geq 4.3$ group and a further 63 with $\chi^2 > 7.82$ and/or $\ltf < 4.3$. Objects with H$\alpha$ excess are marked in Table \ref{table:main_params}.

\subsection{Reddening}

After removing the obvious sub/over-luminous objects and the
emission line stars from the selection we are left with a cleaner selection of 458 $\sim$ non-emission OB candidates and 19 known OB stars available for further examination of their reddening properties.

\begin{figure*}
\begin{center}
\includegraphics[width=0.9\textwidth]{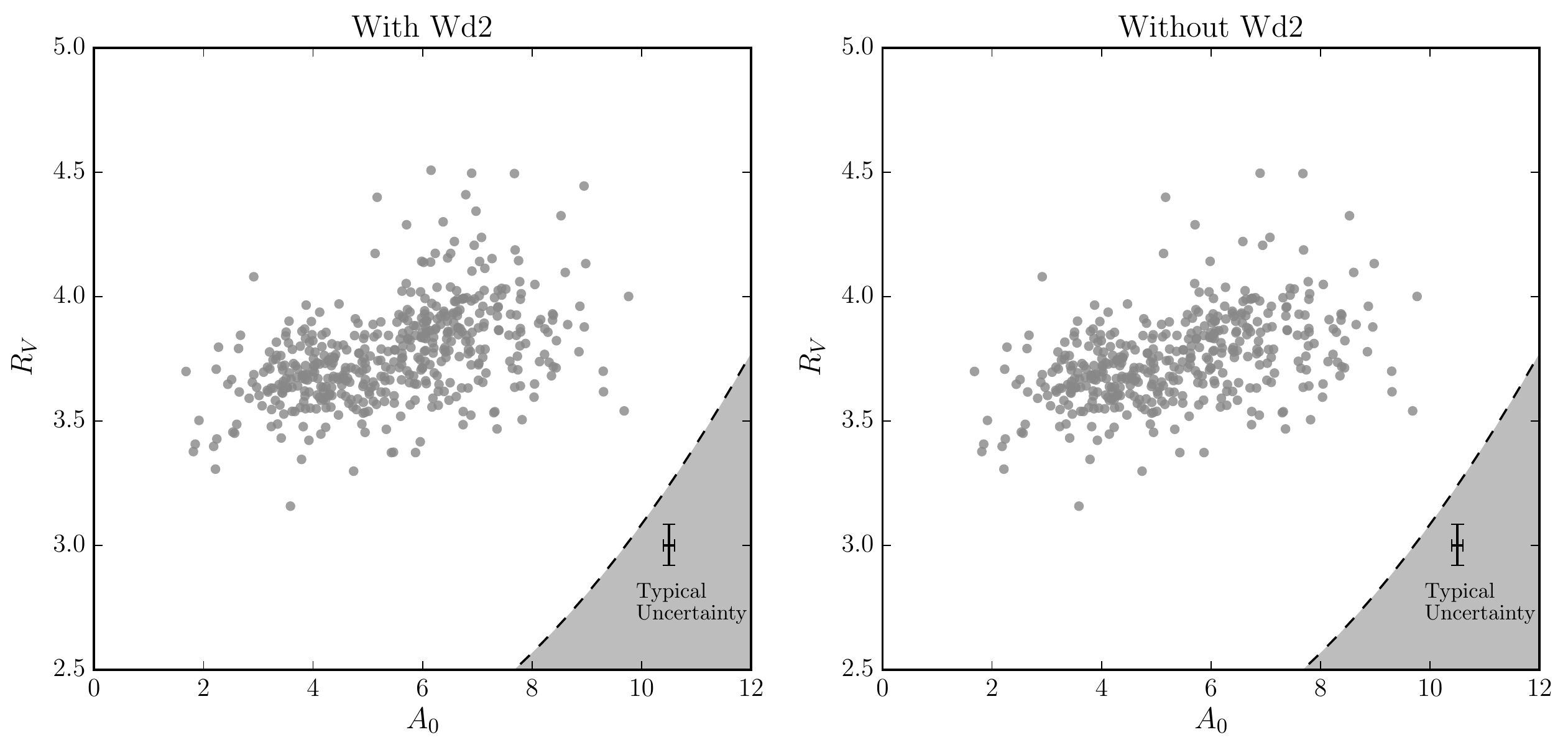}
\caption{$A_0$ vs. $R_V$ plot for the final selection with and without the objects within the 8 arcmin box surrounding Wd 2. There modest increase in $R_V$ as a function of $A_0$ in either case with correlation coefficient $r=0.47$ and $r=0.45$ respectively. The areas shaded in grey are where we
cannot detect OB stars given the survey limits.}
\label{fig:a0rv_plot}
\end{center}
\end{figure*}

Given our tight grasp on $A_0$ and $R_V$, it is of interest to
consider their interdependence.  $R_V$ is plotted as a function of
$A_0$ in Figure \ref{fig:a0rv_plot}. The left panel of this Figure includes those
objects within an 8 arcmin box around Wd 2 and the right hand panel
excludes them. The areas shaded in grey are where we
cannot detect OB stars given the survey limits. In both cases we can see a moderate positive correlation in $R_V$ as a
function of $A_0$ (correlation coefficient $r=0.47$ and $r=0.45$ respectively). On comparing the two panels, it is evident that the members of Wd 2 drive
up the $R_V$ trend more sharply when they are included. The shaded
background shows that the trends seen are independent of the boundaries set
by the survey selection limits. Given that it was demonstrated in Section
\ref{sec:sed_proof} that the fitting method generates negligible
covariance between $A_0$ and $R_V$, we can say with confidence that
the correlation apparent now is related to the physical nature of the
volume of space under study.

It is commonly accepted that increasing $R_V$ is linked to increasing
typical dust grain size, and that values of 3.5 and more are associated
with denser molecular cloud environments
\citep[see e.g.][]{draine2003}.  The $\sim$2 square
degrees under examination here sample sight-lines lying just inside the
Carina Arm tangent direction.  Our pencil beam is evidently one that
would initially pass through the atomic diffuse interstellar medium
and then enter the dense clouds of the Carina Arm, wherein Wd 2 is
located. In this situation it makes sense that as the dust column
grows it becomes ever more dominated by the dense/molecular ISM
component -- i.e. $R_V$ tends to rise.  However the rise is not
dramatic, and the data points show significant dispersion, which may
imply that the variation in the dust properties within the sampled
volume is not especially coherent.  The effect of
the bright limit of the survey is to remove sensitivity to $A_0$
much below 2--3, or to distances less than $\sim$3~kpc (see below).
Current maps of Galactic spiral arm structure place this distance
already within the Carina Arm \citep{Russeil2003, Vallee2014}.

The clear message of Figure \ref{fig:a0rv_plot} is that the typical,
if necessarily idealised, reddening law for this sight-line is $R_V \sim  3.8$,
which rises much less sharply with decreasing wavelength than the Galactic
average of $R_V = 3.1$ \citep[see Figure 13 in][]{fitzpatrickandmassa2007}.

\section{Discussion}

\subsection{The number and spatial distribution of the OB candidates}

\begin{sidewaysfigure*}
\centering
\includegraphics[width=\textwidth]{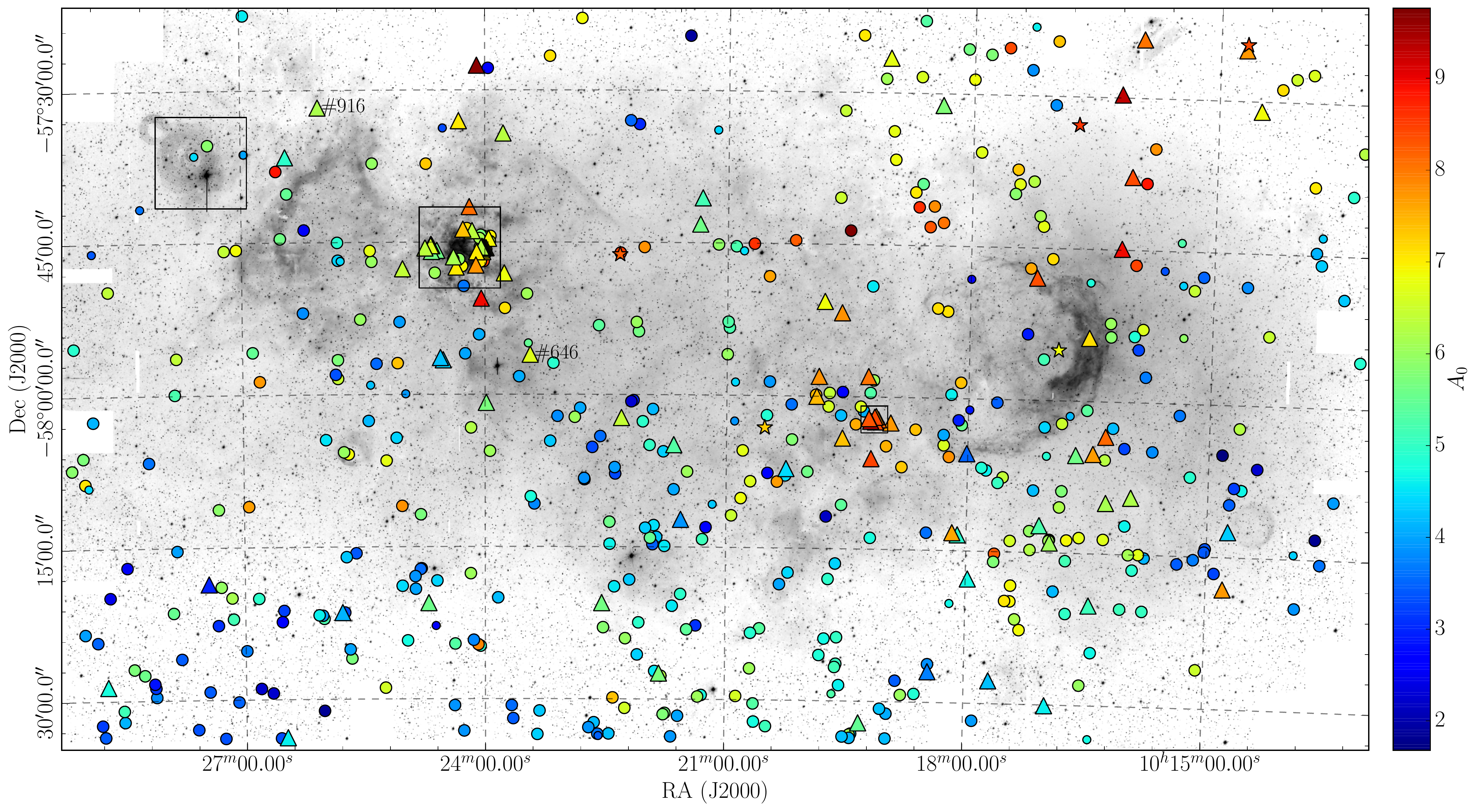}
\caption{Known and candidate OB stars, in our selection, coloured according to their inferred reddening ($A_0$) and over-plotted on the high-confidence inverse VPHAS+ H$\alpha$ image. The boxes pick out the known clusters IC 2581, Wd 2, and DBS2003 45 (working from east to west).  The symbols used have the following interpretations: O stars with $\log(T_{\rm eff}) > 4.477$ are triangles; early B stars, $ 4.300 \leq \log(T_{\rm eff}) \leq 4.477$ are large circles; stars and small circles represent the blue supergiants and under-luminous objects respectively. The white space under the object near the eastern boundary is currently a region of reduced confidence in H$\alpha$ (without inclusion of the adjacent field)
Low resolution spectroscopy of object \#916 and \#646
suggests their spectral types may be early as O3-O4 (Mohr-
Smith et al in prep). \label{fig:area}}
\end{sidewaysfigure*}

Figure \ref{fig:area} shows the location of each new candidate in the 2 square degrees for which the SED fit returned $\chi^2 \leq 7.82$ and $\log(\rm T_{\rm eff}) \geq 4.3$, over-plotted on the VPHAS+ H$\alpha$ mosaic.  Each star is colour-coded according to its derived extinction, $A_0$. The 527 objects are scattered across the field, with lower reddenings ($A_0 < 5$) dominating in the southern half. Apart from in Westerlund 2 itself, the distribution is sparser and more highly-reddened in the north. Towards the NW and the tangent direction, roughly at RA 10h11m, Dec -56 14 (J2000) \citep{Dame2007}, the most reddened objects
($A_0 > 8$) are found.

489 of the objects shown have not been identified previously as confirmed or candidate OB stars. Previous works by \cite{Reed2003}, and by \cite{kaltchevaandgolev2012} have noted a further
26 stars earlier than B3 within this region -- all of which are brighter than $V = 11$, and therefore not in our sample.  Also for reasons of brightness, our sample does not include any stars obviously associated with IC 2581. Turner (1978) studied the cluster -- home to a number of early B stars -- establishing a distance of 2.87~kpc, and a typical reddening corresponding to $A_0 \sim 1.5$.
For the present selection, this cluster is too close and too lowly-reddened: a B3 main sequence star with $A_0 = 1.5$ needs to be at a distance of 3.8 kpc to achieve $g = 13$.  The one star that has been uncovered close to IC~2581 is a candidate sub-luminous object, likely to be at a much shorter distance and unconnected to IC~2581.  It can be seen in Figure \ref{fig:a0rv_plot} that $A_0 = 1.9$ for the least reddened candidate OB star in the sample.

It has been argued before by e.g. \cite{Grabelsky1988} and \cite{Dame2007} that the Carina Arm tangent region traced in CO spans the distance range from 3 to 5~kpc.  At larger distances the conical volume captured here reaches beyond the Solar Circle where declining amounts of molecular gas are detected.  Taking note of these considerations, it seems likely that the high values of $R_V$, trending from 3.6 to 3.9, revealed by our SED fits (Figure \ref{fig:a0rv_plot}) are largely a product of the dominant and increasing contribution to the total extinction from the dust column of the Carina Arm. Similarly \cite{povichetal2011} found it necessary to adopt $R_V=4$ for embedded Carina Arm objects at $l\sim 287^{\circ}$. In contrast, \cite{Turner1978} determined $R_V$ to be 3.11$\pm$0.18 across this region, based on bright OB stars with extinctions below $A_V$ of 2 -- clearly the foreground to our sample. Indeed it seems likely that much of the extinction of the OB population spanning $A_0 \sim 2$ to $A_0 \sim 9$ accumulates within the Carina Arm.  The appearance of Figure \ref{fig:param_plot} indicates few detected main sequence OB stars beyond a distance modulus of 14 (~6kpc).

\subsection{Westerlund 2}

Figures \ref{fig:param_hist} and \ref{fig:a0rv_plot} tell us that a single value of $R_V$ cannot be used to describe the extinction law of sight lines towards all objects in Wd 2. Instead we find that $R_V$ ranges from approximately 3.5 to 4.5 within the cluster. Similar spreads in $R_V$ within star clusters has previously been found by \cite{fitzpatrickandmassa2007} and highlights the importance of deriving $R_V$ on a star-by-star basis. \cite{huretal2015} describe a hybrid extinction model with $R_V = 3.33 \pm0.03$ to $A_0 \sim3$ (based on three stars), while $R_V = 4.14 \pm0.08$ is required for stars in Wd 2. Figure \ref{fig:a0rv_plot} cautions against this clear-cut interpretation, even while our results are numerically consistent with theirs. Reality is more fractal and it is best not to place too much weight on small numbers of stars.

In Figure \ref{fig:param_hist} we noticed a tight distribution in $A_0$ for the objects close to Wd 2 as projected on the sky.
While there are no new OB star candidates in the central region of Wd 2 (within the 8 arcmin box shown in fig \ref{fig:area}), there are a handful of probable O stars scattered across the field that share its extinction. All objects that have extinctions consistent to within 1 $\sigma$ of the mean of known objects in Wd 2 ($5.8 > A_0 > 7.2$) and have $\ltf > 4.477$ are identified in Table \ref{table:runaway}. It is possible that these have been ejected from Wd 2 by dynamical interactions or after supernova explosions in binary systems \citep{allenandpveda197,giesandbolton1986}. Given a derived distance of $\sim5$ kpc to Wd 2, an object that is separated from the cluster by $\sim20$ arcmin on the sky would have to have travelled a minimum distance of $\sim30$ pc in 1--2 Myr. This would equate to a minimum (plane-of-sky) velocity of $\sim25$ km/s. Given that massive stars can attain runaway velocities of up to $\sim200$ km/s through dynamical encounters between binary systems \citep{GGP2010}, it is not unreasonable to consider that these objects have been ejected recently from Wd 2. Alternatively these stars may have formed in situ within the wider star forming region on a similar time scale to the cluster. Low resolution spectroscopy of object \#916 and \#646 suggests their spectral types may be early as O3-O4 (Mohr-Smith et al in prep). Their positions are marked on Figure \ref{fig:area}.

As a final point of interest, we note that the more reddened cluster DBS2003 45, picked out in figure \ref{fig:area} as well, also appears to be surrounded by a scatter of OB stars that are reddened similarly to the cluster.

\begin{table}
\begin{center}

\caption{The reddening parameters and angular separation from the centre of Wd 2 ( RA 10 24 18.5 DEC -57 45 32.3 (J2000)) of new O star candidates with similar reddening to the cluster, outside the 8 arcmin box shown in figure \ref{fig:area}. All objects have $\ltf > 4.477$ and $5.8 > A_0 > 7.2$. See Tables \ref{table:main_photom} and \ref{table:main_params} for the full set of data. \label{table:runaway}}
\begin{tabular}{c c c p{1.2cm}}

\hline
ID&$g$&$A_0$& Separation (arcmin)\\
\hline \noalign{\smallskip}

44&$16.62$&$6.88^{+0.07}_{-0.09}$&$80.21$\\ \noalign{\smallskip}
121&$14.28$&$6.21^{+0.06}_{-0.08}$&$70.59$\\ \noalign{\smallskip}
191&$17.43$&$6.05^{+0.09}_{-0.09}$&$64.98$\\ \noalign{\smallskip}
144&$16.94$&$6.13^{+0.07}_{-0.09}$&$68.46$\\ \noalign{\smallskip}
161&$19.28$&$7.11^{+0.07}_{-0.12}$&$62.73$\\ \noalign{\smallskip}
346&$16.60$&$6.76^{+0.07}_{-0.09}$&$46.51$\\ \noalign{\smallskip}
413&$17.06$&$6.83^{+0.05}_{-0.06}$&$36.42$\\ \noalign{\smallskip}
576&$15.40$&$6.42^{+0.05}_{-0.07}$&$23.13$\\ \noalign{\smallskip}
916&$15.01$&$6.19^{+0.05}_{-0.07}$&$19.67$\\ \noalign{\smallskip}
646&$15.58$&$6.69^{+0.06}_{-0.08}$&$12.60$\\ \noalign{\smallskip}
796&$16.37$&$7.03^{+0.05}_{-0.07}$&$12.50$\\ \noalign{\smallskip}
662&$15.33$&$6.30^{+0.05}_{-0.06}$&$12.06$\\ \noalign{\smallskip}
846&$16.82$&$6.39^{+0.04}_{-0.05}$&$6.03$\\ \noalign{\smallskip}
661&$17.21$&$6.81^{+0.04}_{-0.05}$&$5.05$\\ \noalign{\smallskip}
\hline
\end{tabular}
\end{center}
\end{table}

\subsection{Candidate blue supergiants and sub-luminous stars}

The results from Section \ref{sec:lum} suggest the presence of 5 high luminosity B stars scattered across the field. If they are early-B supergiants,
their absolute visual magnitudes would be in the region of $\sim-6.5$ \citep{crowtheretal2006}. On correcting the previous main-sequence assumption, we find their derived distance moduli, $\mu$, rise from $\sim9$ to $\sim13.5$, placing them amongst the general OB population that we pick out. \cite{MeylanandMaeder1983} estimate a surface density of around 10 - 20 blue supergiants (BSGs) per kpc$^{2}$ in the Galactic Plane. Assuming that our selection spans distances from 2 - 6 kpc we are sampling a projected disk surface area of a little over 1 $\rm kpc^2$; so finding 5 candidates undershoots the surface density prediction but not to the extent that it can be claimed to be inconsistent with it. Given that these candidates are affected by saturation in the $i$ band ($i\lesssim12$), there may one or two BSGs that have fallen into the `poor-fit' group due to saturation in one or more bands.

We also find evidence for the presence of a population of subdwarf stars (see table \ref{table:sub/over-luminous} and figure \ref{fig:param_plot}). Of these 9 may be sdO stars, leaving 23 in the sdB category. The absolute magnitudes of the latter range from $M_V=3-6$ \citep{starkandwade2003}. Since these objects are $\sim6$\,mag fainter than their main-sequence counterparts, their distance moduli are likely to be $\sim10$ as opposed to the estimated values of $\sim16$. This behaviour and the spatial scattering of the subdwarf candidates suggests that we are looking at a group of moderately reddened $A_0 \sim4$ stars in the foreground of the main OB population. We are biased to select more highly reddened subdwarf stars due to the 2MASS faint limit as discussed in Section \ref{summaryofresults}. If this limit was not in place we would expect to find more lowly reddened sdB stars in the selection.

Although the SED-fitting we have performed has no sensitivity to
surface gravities and limited sensitivity to stellar effective
temperature, the fact that the Carina Arm region studied falls near the
tangent has allowed us to pick out the extreme objects purely from
their outlying distance moduli -- relative to the near MS stars
concentrated in the range $11 < \mu < 14$. While this approach works
here, it is evident that in other sight-lines, where the population of
OB stars may be spread more uniformly across a larger distance range,
the luminosity extremes would not stand out in the same way.

\section{Conclusions}

In summary, we have demonstrated a method for selecting and
parametrizing the reddening and basic stellar properties of OB stars
uniformly across large areas of the Southern sky using VPHAS+, and
NIR survey data.  The selection presented here has resulted in reddening parameters for 848 O and B stars (see table \ref{table:breakdown}). Of these, 489 are well-fit new OB stars hotter than 20000 K, including 74 probable O stars. This has been achieved by reaching down to $g
= 20$\,mag and approaches a factor of 10 increase relative to the small number of known and candidate O to B2 stars in the region.

By bringing together VPHAS+ $u,g,r,i$ photometry with NIR 2MASS
photometry, we are able to determine both the value of the extinction,
$A_0$, and test and select the most appropriate reddening law, as
parametrised by $R_V$, to a high degree of accuracy: both are
typically measured to better than 0.1 (magnitudes in the case of $A_0$).
Pleasingly there are signs that the still preliminary nature of the
photometric calibration of the VPHAS+ survey data blends well with the now well-established 2MASS calibration.

We set out expecting to only gain a crude impression of stellar
effective temperatures (and hence distance moduli), and so it has
turned out.  But we have found a satisfying consistency with earlier
results in our benchmark region around the much-studied cluster
Westerlund 2, confirming that our methods are sound and able to e.g
distinguish early O stars from late-O and early-B stars.  This
represents an efficient start to selection that needs to be followed
up by spectroscopic confirmation and measurement of stellar
parameters.  With precise spectroscopic parameters in hand, the
photometry can be re-used for direct and even more precise measurement of reddening laws.

We have also seen how the high resolution and wide field of view of
OmegaCam can bring a wider context to the study of open clusters and
OB associations, through an ability to identify potentially-related
stars that have either been ejected from clusters or simply have
formed -- perhaps as part of a wider star-formation event -- in
relative isolation in the surrounding field. This study has also uncovered 5 BSG candidates as well as 32 reddened candidate subdwarfs of which 9 may be sdO stars.

In the future, we aim to roll out this method to support the complete
characterisation of the massive-star population and the patterns
of extinction they can reveal across the entire Southern Galactic
mid-plane to distances of $\sim5$kpc or more. \cite{garmanyetal1982} were able to claim a volume-limited census to $\sim2.5$kpc 3 decades ago
-- now it should be possible to expand the effective volume by a
factor of 4 or so, with the difference this time that Gaia parallaxes
as they appear will bestow a confidence as to what the volume limits actually are.

\section*{Acknowledgements}

This paper is based on data products from observations made with ESO Telescopes at the La Silla Paranal Observatory under programme ID 177.D-3023, as part of the VST Photometric Hα Survey of the Southern Galactic Plane and Bulge (VPHAS+, www.vphas.eu). Use is also made of data products from the Two Micron All Sky Survey, which is a joint project of the University of Massachusetts and the Infrared Processing and Analysis Center/California Institute of Technology, funded by the National Aeronautics and Space Administration and the National Science Foundation.

This research made use of Astropy, a community-developed core Python package for Astronomy \citep{crawfordetal2013} and TopCat \citep{Taylor2005}.

The authors wish to thank the referee for helpful comments that have improved this paper. MM-S acknowledges a studentship funded by the Science and Technology Facilities Council (STFC) of the United Kingdom (ref.
ST/K502029/1). JED and GB acknowledge the support of a research grant funded by the STFC (ref. ST/J001333/1). NJW is in receipt of a fellowship funded by the Royal Astronomical Society of the United Kingdom.

\footnotesize{
\bibliographystyle{mn2e}
\bibliography{wd2}
}

\end{document}